\documentclass[preprint2]{aastex}
\usepackage{amsbsy}

\def\i{\hbox{\,{\sc i}}}
\def\ii{\hbox{\,{\sc ii}}}

\def\iv{\hbox{\,{\sc iv}}}

\newcommand{\teff}{\mbox{$T_{\mathrm{eff}}$}}

\shorttitle{A study of the $H\alpha$ spectrum of TT Hya}
\shortauthors{Budaj, Richards \& Miller}

\begin{document}


\title{A Study of Synthetic and Observed $H\alpha$ Spectra of TT Hydrae}


\author{
J\'{a}n Budaj\altaffilmark{1},
Mercedes T. Richards
and Brendan Miller}
\affil{Department of Astronomy \& Astrophysics, Penn State University,
525 Davey Laboratory, University Park, PA, 16802, USA}
\email{budaj@astro.psu.edu, mtr@astro.psu.edu,
bmiller@astro.psu.edu}

\altaffiltext{1}{Astronomical Institute, Tatransk\'{a} Lomnica,
05960, Slovak Republic, http://www.ta3.sk/$\sim$budaj}

\begin{abstract}
The formation and properties of accretion discs and circumstellar material
in Algol-type systems is not very well understood. 
In order to study the underlying physics of these structures, 
we have calculated synthetic $H\alpha$ spectra of TT Hya, which
is an Algol-type eclipsing binary with an accretion disc.
Both the primary and secondary stars were considered in 
the calculations as well as a disc surrounding the primary.
The Roche model for the secondary star was assumed.
The synthetic spectra cover all the phases including primary eclipse
and are compared with the observed spectra.
The influence of various effects and free parameters of the disc
on the emerging spectrum was studied.
This enabled us to put some constraints on the geometry,
temperature, density and velocity fields within the disc.
Differences found between the observed and synthetic spectra
unravel the existence of a gas stream as well as a hotter disc-gas
interaction region. An additional cooler circumstellar region
between the C1 and C2 Roche surfaces is suggested to account for various
observed effects. 
A new computer code called {\sc{shellspec}} was created for this purpose
and is briefly described in this work as well.
It is designed to solve simple radiative transfer along the line
of sight in 3D moving media. The scattered light from a central 
object is taken into account assuming an optically thin  
environment. Output intensities are then integrated through  
the 2D projection surface of a 3D object. 
The assumptions of the code include LTE and optional known state
quantities and velocity fields in 3D. 
\end{abstract}


\section{Introduction}
TT Hydrae (HD 97528, HIP 54807, SAO 179648, 
$V=7.27^{m}, \alpha=11^{h}13^{m}, \delta=-26^{\circ}28'$) 
is a typical Algol-type eclipsing binary system with an orbital period
of $P_{\rm orb}=6.95$ days (Kulkarni \& Abhyankar 1980) and consists
of a hotter B9.5 V main sequence primary and a cooler evolved K1 III-IV
secondary star filling its Roche lobe. There is 
evidence of the circumstellar material in the form of an accretion 
disc (Plavec \& Polidan 1976) surrounding the primary as well as
an indication of a gas stream (Peters \& Polidan 1998).
Kulkarni \& Abhyankar (1980) obtained UBV photometric observations
and photometric elements of the system. These observations were re-examined
by Etzel (1988) who found from photometry, spectrophotometry and 
spectroscopy:
$\teff=9800$K, $v\sin i=168\pm5$ km\,s$^{-1}$ for the primary, and 
$\teff=4670-4850$K for the secondary. He also suggested, 
based on the Inglis-Teller formula, that the gas producing the UV excess 
and Balmer line emission has an electron number density lower 
than $10^{12}$ cm$^{-3}$. He also estimated a distance of 193 pc to 
the binary.
Eaton \& Henry (1992) estimated $v\sin i=43\pm3$ km\,s$^{-1}$ for the
secondary.
Plavec (1988) studied the IUE spectra of TT Hya and concluded from 
the presence of Fe\ii\ absorption lines that the vertical dimension of 
the accretion disc is significant and is at least comparable to 
the diameter of the primary.
He estimated a distance to the system of 194 pc, and pointed out that
the presence of super-ionized emission lines of Si\iv, C\iv, N\,{\sc v}
poses a question about the source of ionization.
Peters (1989) estimated from the eclipses of the $H\alpha$ emission 
region that a fairly symmetrical disc fills up to 95\% of 
the Roche lobe radius. Moreover, she argued that the disc must be
rather flat with the vertical dimension comparable or less than 
the diameter of the primary. The mean electron number density in the disc 
was estimated to be $\sim 10^{10}$ cm$^{-3}$. 
The depth of the $H\alpha$ absorption was found to be strongly variable
with phase and was deepest shortly before and after primary eclipse;
a feature also observed in other Algols (e.g., Richards 1993). 
Peters (1989) also suggested that the additional absorption in the center
of the $H\alpha$ line probably comes from 
the dense inner region of the accretion disc. She also observed that
the $H\alpha$ core is blueshifted relative to the primary by about 
$50$ km\,s$^{-1}$ and interpreted it as evidence of the mass outflow.

Sahade \& Cesco (1946) determined the orbital parameters of the primary
and identified an eccentric orbit from the Ca\ii\,K and H lines.
%Miller \& McNamara (1963) measured the radial velocities
%of the primary and secondary from a few spectra.  
%In a more extensive study, 
Popper (1989) measured radial velocities of 
the primary and secondary and found that the motion of the secondary
was consistent with the circular orbit 
($K=132$ km\,s$^{-1}$) but not in the case of the primary.
Van Hamme \& Wilson (1993) reanalyzed the light and velocity curves 
separately and simultaneously and used a physical model to obtain
the following parameters:
separation $a=22.63\pm0.12 R_{\odot}$, effective temperature of the
primary $T_{1}=9800$K, secondary $T_{2}=4361$K, radius of the primary 
$R_{1}=1.95 R_{\odot}$, secondary $R_{2}=5.87 R_{\odot}$,
mass of the primary $M_{1}=2.63M_{\odot}$, secondary $M_{2}=0.59M_{\odot}$, 
and mass ratio $q=0.2261\pm0.0008$.
Albright \& Richards (1996) obtained the first Doppler tomogram
to reveal an indirect image of the accretion disc.
Hipparcos (ESA 1997) obtained a parallax of $6.50\pm 0.95$ mas.
Peters \& Polidan (1998) observed red-shifted absorption in N\i, N\ii\ lines 
in their FUV spectra at the phase 0.95 and interpreted it as evidence
of the gas stream and inferred the rate of inflow to be greater than
$10^{-12}M_{\odot}$yr$^{-1}$. Most recently, Richards \& Albright (1999) 
studied the $H\alpha$ difference profiles
and properties of the accretion region and put TT Hya into the context
of other Algol binaries.

It would be interesting to verify the above estimates,
conclusions, and interpretations on more physical grounds using some 
more sophisticated methods involving synthetic spectra  
which would take into account the circumstellar matter.
Unfortunately, there are not many tools available for such a task at present.

There are sophisticated computer codes for calculating and inverting
light curves or spectra of binary stars with various shapes
or geometry including the Roche model 
(Lucy 1968; Wilson \& Devinney 1971; Mochnacki \& Doughty 1972;
Rucinski 1973; Hill 1979; Zhang et al. 1986; Djurasevic 1992;  
Drechsel et al. 1994; Vink\'{o} et al. 1996; Hadrava 1997;     
Bradstreet \& Steelman 2002; Pribulla 2004).
In these codes, the stars are assumed to be nontransparent and
stripped of any circumstellar matter.
An interesting approach to study extended semitransparent atmospheres 
of some eclipsing binary components was developed by
Cherepashchuk et al. (1984). 
Often the 3D model of the circumstellar matter
(behavior of state quantities and velocity field)
is known or expected as a result of hydrodynamic simulations or 
observational constraints (see for example Karetnikov et al. 1995;
Richards \& Ratliff 1998).
Unfortunately, 3D Non-LTE (NLTE) calculations and spectral synthesis
including complex hydrodynamics are difficult to carry out so 
one alternative has been to perform a simple volume integration 
of emissivity, which is often too oversimplified for 
the particular problem.

On the other hand, highly sophisticated model atmospheres and
spectrum synthesis codes have been developed assuming NLTE 
and plane-parallel atmospheres of hot stars
(Hubeny 1988; Hubeny \& Lanz 1992, 1995; Hubeny et al. 1994),
spherically symmetric atmospheres (Kub\'{a}t 2001), 
or stellar winds (Krti\v{c}ka \& Kub\'{a}t 2002).   
There are also sophisticated stationary pla\-ne-pa\-ra\-llel
li\-ne-blan\-ke\-ted model atmospheres and spectrum synthesis codes 
for a large variety of stars assuming LTE 
(Kurucz 1993a, 1993b; Smith \& Dworetsky 1988; Piskunov 1992; and many others).
However, these are very specialized codes and their main purpose
is to calculate the spectrum emerging from a stellar atmosphere, and it is   
difficult to apply them to the various cases outlined above.
An exception is the special case of circumstellar matter in the form of
accretion discs in cataclysmic variables (CVs). 
In this case, the disc is either approximated by a set of geometrically
thin, but optically thick, static local atmospheres and the output radiation
is a sum of properly Doppler shifted local emerging intensities 
(Orosz \& Wade 2003; Wade \& Hubeny 1998; la Dous 1989) or, in case of
optically thin discs or accretion disc winds, the Sobolev approximation
is used (Proga et al. 2002; Long \& Knigge 2002; Rybicki \& Hummer 1983).
Linnell \& Hubeny (1996) developed a package of codes which can calculate
light curves or spectra of interacting binary stars including optically
thick (nontransparent) disc.
However, it is often necessary to solve the radiative transfer in a moving
disc at least along the line of sight as demonstrated by
Horne \& Marsh (1986). Nevertheless, their calculations assumed a linear 
shear and neglected stimulated emission, continuum opacity, scattering, 
as well as a central star.
 
In this paper, we attempt to bridge the gap in the previously mentioned 
approaches using our new code {\sc{shellspec}}. It is a tool which 
solves in LTE the simple radiative transfer along the line of sight 
in an optional optically thin 3D moving medium.
Optional transparent (or non transparent) objects such as a spot, disc, 
stream, jet, shell or stars  as well as an empty space may be defined  
(or embedded) in 3D and their composite synthetic spectrum 
calculated. The stars may have the Roche geometry and   
known intrinsic spectra.
The code is quite a multi-purpose, independent, and flexible tool which can 
also calculate a light curve or a trailing spectrogram where necessary, 
or can be used to study various objects or effects.
In Section \ref{s1}, we describe the basic astrophysics included 
in the code.
In Section \ref{snp}, we briefly mention numerical methods used in the code.
In Section \ref{s2}, we study in more detail 
the effects of various input parameters on the spectrum of a star 
which has a circumstellar disc.
Budaj \& Richards (2004) made preliminary calculations for a single star 
with a disc with parameters similar to TT Hya. However, these were based on 
the stellar parameters found by Etzel (1988), a wedge shape geometry 
of the disc with a relatively large inner radius (2.5$R_{\odot}$), and 
no limb darkening of the primary star. 
More realistic stellar parameters by Van Hamme \& Wilson (1993) and a slab
shaped disc geometry will be used in this work.
Then in Section \ref{s3}, we will make more detailed calculations
of the system for different phases involving both stars and a disc 
and will compare our new synthetic spectra to real observations of TT Hya.

\section{Basic Astrophysics}
\label{s1}
This section describes the basic astrophysics included
in the code. For more detailed information and a manual for the code,
the reader is referred to Budaj \& Richards (2004).

\subsection{Radiative Transfer}
In the following analysis, the calculations are carried out in
the observer's Cartesian frame with $z$ pointing towards the observer.
The radiative transfer equation along the line of sight is:
\begin {equation}
dI_{\nu}=(\epsilon_{\nu}-\chi_{\nu}I_{\nu})dz
%\label{e1}
\end {equation}
where $I_{\nu}$ is the specific monochromatic intensity at
the frequency $\nu$, 
$\chi_{\nu}$ is the opacity,
$\epsilon_{\nu}$ is the emissivity and
$z$ is the distance along the beam.   
%We define the optical depth along the beam
%$\tau_{\nu}=\int \chi_{\nu}dz$.
It is convenient to split the opacity into two contributions,
the true absorption $\kappa_{\nu}$ and the scattering $\sigma_{\nu}$:
\begin {equation}
\chi_{\nu}=\kappa_{\nu}+\sigma_{\nu}.
\end {equation}

Assuming LTE, the line opacity corrected for a stimulated emission
is stated simply as:
\begin {equation}   
\chi_{\nu}^{line}=(1-e^{-\frac{h\nu}{kT}})
N_{l}B_{lu}h\nu\varphi_{lu}(\nu-\nu_{0})(4\pi)^{-1}
%\label{e1}
\end {equation}
where $h$ is the Planck constant,
$h\nu$ is the energy of the transition from the lower level $l$ to
the upper level $u$,
$k$ is the Boltzmann constant,
$T$ is the temperature,
$N_{l}$ is the population of the $l$-th state of the corresponding ion, and
$B_{lu}$ is the Einstein coefficient for the whole solid angle.
The velocity field enters the equation via the shifted normalized Voigt
profile $\varphi_{lu}(\nu-\nu_{0})$ where
\begin {equation}
\nu_{0}=\nu_{lu} \left( 1 + \frac{v_{z}(z)}{c} \right)
\end {equation}
where $\nu_{lu}$ is the laboratory frequency of the line and
$v_{z}={\boldsymbol v}.{\boldsymbol n}$ is the radial velocity (positive 
towards the observer) or projection of the local velocity vector 
${\boldsymbol v}$ to the line of sight unit vector ${\boldsymbol n}$.
The Einstein coefficient, $B_{lu}$, is related to the oscillator
strength, $f_{lu}$, by:
\begin {equation}
B_{lu}=\frac{4\pi^{2} e^{2} f_{lu}}{m_{\rm e}c h\nu_{lu}}
%\label{e2}
\end {equation}
where $e, m_{\rm e}$ are the electron charge and mass, respectively and
$c$ is the speed of light. 
%$g_{l,i}, g_{u,i}$ are statistical weights of the lower and upper levels.
The shape of the Voigt profile is determined by the thermal
and the microturbulent broadening, $v_{trb}$,
characterized by the Doppler half-width
\begin {equation}
\Delta\nu_{D}=\frac{\nu}{c}\sqrt{\frac{2kT}{m}+v^{2}_{trb}}
\end {equation}
as well as by the damping broadening characterized by
%$u=\Delta\nu/\Delta\nu_{D}$,
the frame damping parameter  
\begin {equation}
a=\gamma/(4\pi\Delta\nu_{D})
\end {equation}
where the damping constant
\begin {equation}
\gamma=\gamma_{Nat.}+\gamma_{Stark}+\gamma_{VDW}
\end {equation}
includes the contribution from the Natural, Stark and Van der Waals
broadening. 
In the case of LTE, all the line opacity 
is due to the true absorption process i.e.:
\begin {equation}
\kappa_{\nu}^{line}=\chi_{\nu}^{line}
\end {equation}
We also included four other continuum opacity sources:
the HI bound-free opacity, the HI free-free opacity,  
Thomson scattering and Rayleigh scattering on neutral hydrogen,
denoted by $\kappa_{\nu}^{HIbf}, \kappa_{\nu}^{HIff},
\sigma_{\nu}^{TS}, \sigma_{\nu}^{RS}$, respectively.\\
The total true absorption $\kappa_{\nu}$ is the sum of the three opacity
sources:
\begin {equation}
\kappa_{\nu}=\kappa_{\nu}^{line}+\kappa_{\nu}^{HIbf}+\kappa_{\nu}^{HIff}.
\end {equation}
The total scattering $\sigma_{\nu}$ is:
\begin {equation}
\sigma_{\nu}=\sigma_{\nu}^{TS}+\sigma_{\nu}^{RS}.
\end {equation}

The thermal emissivity associated with the true absorption can then be
written as 
\begin {equation}
\epsilon_{\nu}^{th}=B_{\nu}(T(z))\kappa_{\nu}
\end {equation}
where $B_{\nu}$ is the Planck function for the local value of the
temperature.
For scattering emissivity we have
\begin {equation}
\epsilon_{\nu}^{sc}=
\int\!\!\int \sigma(\nu',\nu,{\boldsymbol n'},{\boldsymbol n})
I(\nu',{\boldsymbol n'})d\nu'\frac{d\omega'}{4\pi}
\end {equation}
where $\sigma(\nu',\nu,{\boldsymbol n'},{\boldsymbol n})$ is the scattering 
coefficient containing the general redistribution function.
It is this term which
causes the main difficulty, since apart from redistributing the frequencies
($\nu' \rightarrow \nu$), it also couples the radiation in one direction   
${\boldsymbol n}$ with the radiation field in all other directions 
${\boldsymbol n'}$.  However, in many applications (e.g., optically thin 
shells) this term can either be neglected or governed by the scattering 
of light from the central object. 
We assume coherent isotropic scattering (as seen
from the scattering particle frame) from 
a blackbody or from a central spherical star with precalculated surface
intensity $I^{\star}_{\nu}$ or flux $F^{\star}_{\nu}$. In this case    
the emissivity reduces to:
\begin {equation}
\epsilon_{\nu}^{sc}=\sigma_{\nu}J_{\nu}
\end {equation}
where $J_{\nu}$ is the mean intensity. Ignoring limb darkening, $J_{\nu}$
can be approximated by:
\begin {equation}
J_{\nu} \approx I^{\star}_{\nu_{1}}\omega/4\pi
\end {equation}
where $\omega$ is the solid angle subtended by the central star and
\begin {equation}
\omega/4\pi=\frac{1}{2}
\left( 1-\sqrt{1-\left(\frac{R_{\star}}{r}\right)^{2}} \right)
\end {equation}
where $R_{\star}$ is the radius of the central star and
$r$ is the distance from the center (of the star/grid) and
\begin {equation}
\nu_{1}=\nu \left( 1 - \frac{v_{1}}{c} \right)
\end {equation}
and
\begin {equation}
v_{1}=-\frac{{\boldsymbol r}.({\boldsymbol v}-{\boldsymbol v}_{\star})}{r}
+v_{z}
\end {equation}
where ${\boldsymbol v}$ is the velocity field vector at the given point 
specified by the vector ${\boldsymbol r}$ and ${\boldsymbol v}_{\star}$ 
is the velocity of the center of mass of the central object.\\
For $R_{\star}/r<<1$, an approximation including the limb
darkening and the non-isotropic dipole phase function 
$g({\boldsymbol n'},{\boldsymbol n})=
\frac{3}{4}(1+({\boldsymbol n'}.{\boldsymbol n})^{2})$ can be used:
\begin {equation}
\epsilon_{\nu}^{sc}=\frac{3}{4}\left(1+\frac{r_{z}^{2}}{r^{2}}\right)
\sigma_{\nu}J_{\nu}
\end {equation}
where
\begin {equation}
J_{\nu} \approx \frac{F^{\star}_{\nu_{1}}}{4\pi} \frac{R_{\star}^{2}}{r^{2}}
\end {equation}
where
\begin {equation}
F^{\star}_{\nu_{1}}= \pi I^{\star}_{\nu_{1}} \left(1-\frac{u}{3}\right)
\label{e3}
\end {equation}
where $u$ is the limb darkening coefficient.\\
The total emissivity is then
\begin {equation}
\epsilon_{\nu}=\epsilon_{\nu}^{th}+\epsilon_{\nu}^{sc}
\end {equation}
and the total source function is:
\begin {equation}
S_{\nu}=\epsilon_{\nu}/\chi_{\nu}.
\end {equation}

The radiative transfer equation along the line of sight is solved
(see Section \ref{snp}) and the flux, $F_{\nu}$, from the object at 
the Earth is then obtained by the integration of the output intensities 
through the 2D projection surface of the 3D object.

\subsection{Roche Geometry}
\label{srg}
Both objects, star and companion, may have shapes according to the Roche
model for detached or contact systems. Descriptions of the Roche
model can be found in Kopal (1959), Limber (1963), 
Plavec \& Kratochvil (1964), Mochnacki \& Doughty (1972), Hilditch (2001)
and many other papers and books.
Most of them use cylindrical coordinates but we will assume 
a Cartesian coordinate system $(x,y,z)$ centered on one
of the stars (labeled as 1) such that the companion (labeled as 2) is
at (1,0,0) and revolves around the $z$ axis in the direction of 
the positive $y$ axis.
Let the mass ratio, $q$, always be $m_{2}/m_{1}$ or
`companion/star' and $q<1$ will indicate the central star is heavier 
while $q>1$ means the central star is lighter. 
We will also assume synchronous rotation i.e., the stellar 
surface rotates with the orbital period.
Then, the normalized Roche potential, $C(x,y,z)$, is expressed as:
\begin {equation}                                        
C=\frac{2}{(1+q)r_{1}}+\frac{2q}{(1+q)r_{2}}+     
\left(x-\frac{q}{1+q} \right)^{2}+y^{2}
\end {equation}
where 
\begin {equation}
r_{1}=\sqrt{x^{2}+y^{2}+z^{2}}, 
\end {equation}
and
\begin {equation}
r_{2}=\sqrt{(x-1)^{2}+y^{2}+z^{2}}.
\end {equation}
The Roche surface of a detached component is defined as an equipotential
surface $C_{s}=C(x_{s},y_{s},z_{s})$ passing through the sub-stellar point
$(x_{s},y_{s},z_{s})$ (point on the surface of the star in between the
stars,
$0<x_{s}<1, y_{s}=z_{s}=0$) which is 
localized by the `fill-in' parameter $f_{i}\leq 1$.
We define this by:
\begin {equation}                                     
f_{i}=x_{s}/L_{1x}, ~~~~~~~f_{i}=(1-x_{s})/(1-L_{1x})
\end {equation}
for the primary and the secondary, respectively.
$L_{1x}$ is the $x$ coordinate of the L1 point at ($L_{1x},0,0$).
The Roche equipotential surface $C_{s}$ of a contact system will be defined
by the fill-out parameter $1< f_{o}\leq 2$:
\begin {equation}
f_{o}=\frac{C1-C_{s}}{C1-C2}+1
\end {equation}  
where potentials $C1,C2$ correspond to the potentials at the L1 and L2
points, respectively.
First, we calculate L1, L2, $C_{s}$ and $x$-boundaries of the object
using the Newton-Raphson iteration method in the $x$ direction,
and then the 3D shape of the surface is solved using the Newton-Raphson
iteration in $y$ and $z$ coordinates with the precision of about $10^{-5}$.

Gravity darkening is taken into account by varying the surface
temperature according to the following law (Claret 1998):
\begin {equation}
T/T_{p}=(g/g_{p})^{\beta}
\end {equation}  
where $g$ is the normalized surface gravity, $\beta$ is the gravity
darkening exponent, 
$T_{p}$, and $g_{p}$ are the temperature and gravity at the rotation pole.
The normalized gravity vector is ${\boldsymbol g}=(C_{x},C_{y},C_{z})$ and
\begin {equation}
g=\sqrt{C_{x}^{2}+C_{y}^{2}+C_{z}^{2}},
\end {equation}  
where:
\begin {equation}
C_{x}=\frac{\partial C}{\partial x}, 
C_{y}=\frac{\partial C}{\partial y},
C_{z}=\frac{\partial C}{\partial z}.
\end {equation}  
The gravity darkening factor of the surface intensity is then calculated as:
\begin {equation}                                               
f_{GD}=B_{\nu}(T)/B_{\nu}(T_{p}).                               
\end {equation}  
Notice, that there is an imminent singularity in the calculations
in the vicinity of L1, L2 points since gravity falls to zero 
which drags temperatures (a denominator in many equations) to zero.
We avoid the problem by setting the lowest possible value of 
$g/g_{p}=10^{-4}$.
Limb darkening is taken into account using Eq. \ref{eld} and by calculating
the cosine of the angle $\theta$ between the line of sight unit vector 
${\boldsymbol n}=(n_{x},n_{y},n_{z})$ and a normal to the surface:
\begin {equation}
\cos \theta=-{\boldsymbol n}.{\boldsymbol g}/g
=-\frac{n_{x}C_{x}+n_{y}C_{y}+n_{z}C_{z}}
{\sqrt{C_{x}^{2}+C_{y}^{2}+C_{z}^{2}}}.  
\end {equation}  
The reflection effect
%of objects subject to Roche geometry
is not included in the present version.

\section{Numerical Performance}
\label{snp}

\subsection{Solving the Radiative Transfer Equation}
A number of optional objects (including transparent, nontransparent,
empty space) can be defined within the model and the line of
sight may cross more of them within a few grid points.  
A simple and stable method is needed to cope with such  
velocity, density, temperature fields which are optional
and are allowed to be non-continuous.      
Our problem is that of the integration of a first order ordinary
differential equation with known initial values at one boundary.
The equation of the radiative transfer along the line of sight  
at the frequency $\nu$ is integrated with an analogue
of the mid-point method (Budaj \& Richards 2004).
It can be written in the discretised form:  
\begin {equation}
\frac{I_{i+1}-I_{i}}{z_{i+1}-z_{i}}=
\epsilon_{i+1/2}-\chi_{i+1/2}(I_{i+1}+I_{i})/2
\end {equation}
and simply integrated via the following recurrent formula:
\begin {equation}
I_{i+1}=\frac{S_{i+1/2}}{A+1/2}+I_{i}\frac{A-1/2}{A+1/2}
\end {equation}
where
\begin {equation}
A=\frac{1}{\chi_{i+1/2}(z_{i+1}-z_{i})}
\end {equation}
and
\begin {equation}
S_{i+1/2}=\frac{\epsilon_{i+1/2}}{\chi_{i+1/2}},~~~~ I_{1}=0
\end {equation}
where
\begin {equation}
\epsilon_{i+1/2}=\frac{\epsilon_{i}+\epsilon_{i+1}}{2}~~~,~~~~~
 \chi_{i+1/2}=\frac{\chi_{i}+\chi_{i+1}}{2} .
\end {equation}

If the line of sight happens to hit a nontransparent object,
the integration starts on the other side
of the object with the boundary condition:
\begin {equation}
I_{1}=I^{\star}(\nu_{2}) f_{LD}f_{GD}
\end {equation}
or
\begin {equation}
I_{1}=B_{\nu_{2}}(T_{\rm eff}) f_{LD}f_{GD}
\end {equation}
where $\nu_{2}$ is the Doppler shifted frequency in the comoving frame
corresponding to the frequency $\nu$ of the observer's frame:
\begin {equation}
\nu_{2}=\nu \left( 1 - \frac{v^{\star}_{z}}{c} \right)
\end {equation}
and $v^{\star}_{z}$ is the radial velocity of the surface of the
nontransparent object where it intersects the line of sight.
$I^{\star}$ is the surface intensity of the nontransparent object
perpendicular to the surface in the comoving (frozen to the surface) frame. 
If $I^{\star}$ is not available it can be calculated 
e.g., using Eq.~\ref{e3} from the known surface flux. 
Rotation of the nontransparent objects is fully taken into account here
by including it into the calculations of $v^{\star}_{z}$.
$f_{LD}$ is a limb darkening factor:
\begin {equation}
f_{LD}=1-u+u \cos \theta
\label{eld}
\end {equation}
where $u$ is the limb darkening coefficient and $\theta$ is the angle
between the normal to the surface of the nontransparent object and 
the line of sight. $f_{GD}$ is a gravity darkening factor which is 
important in the case of Roche geometry (see the Section \ref{srg})  
and in which case $\teff=T_{p}$ is the temperature at the rotation pole of
a detached star or the temperature at the rotation pole of the more massive 
star in the case of a contact system, otherwise, it is the common effective 
temperature of the spherical star.
If the line of sight happens to pass through an empty space
this region is skipped and the integration continues with
$I_{i+1}=I_{i}$.

The {\sc{shellspec}} code enables the user to look on the object from
different points of view and to calculate the corresponding spectra. 
The input model of the shell is defined in its `body frozen' Cartesian
coordinates ($x'', y'', z''$) with the $z''$ axis corresponding to
the intrinsic rotation axis of the model. 
The spectrum is always calculated in the observer's `line of sight'
Cartesian frame ($x, y, z$) with $z$ pointing to the observer and  
which has the same center of coordinates.

\subsection{Additional Information and Adopted Routines}
The code assumes the known behavior of state quantities: temperature $T$,
density $\rho$, and electron number density $n_{e}$. 
Solar abundances are assumed (Grevesse \& Sauval 1998) but the user
is allowed to change all the element abundances. 
The level populations are obtained from the Boltzmann and Saha equations.
Partition functions (routine `pfdwor') were taken from 
the {\sc{uclsyn}} code (Smith \& Dworetsky 1988; Smith 1992).
A {\sc{fortran77}} code containing the partition function routines
is also available in Budaj, Dworetsky \& Smalley (2002). 
The Gaunt factors (routines `gaunt', `gfree') and Voigt profile
(routine `voigt0') are calculated with 
the subroutines taken from the {\sc{synspec}} code (Hubeny et al. 1994). 
%Damping constants can be found from the VALD
%atomic line database (Kupka et al. 1999) or in Kurucz (1993a).
If the damping constants are not known they are estimated in the code
in the way analogous to the {\sc{synspec}} code.
We also adopted the atomic data for chemical elements (routine `state0')
from {\sc{synspec}} and routines `locate', `hunt' from Numerical Recipes
(Press et al. 1986).
We also used a few sections from our previous original codes for
calculations of radiative accelerations in stellar atmospheres of
hot stars (Budaj \& Dworetsky 2002). Apart from the above, the code was
written from scratch and provides an independent tool to study
a large variety of objects and effects.
CGS units are used within the code and the manuscript, 
if not specified otherwise.

\section{Application to Accretion Disc in TT Hya Type Systems}

%\clearpage
\begin{figure*}
\centerline{
\includegraphics[width=6.cm,angle=-90,clip=]{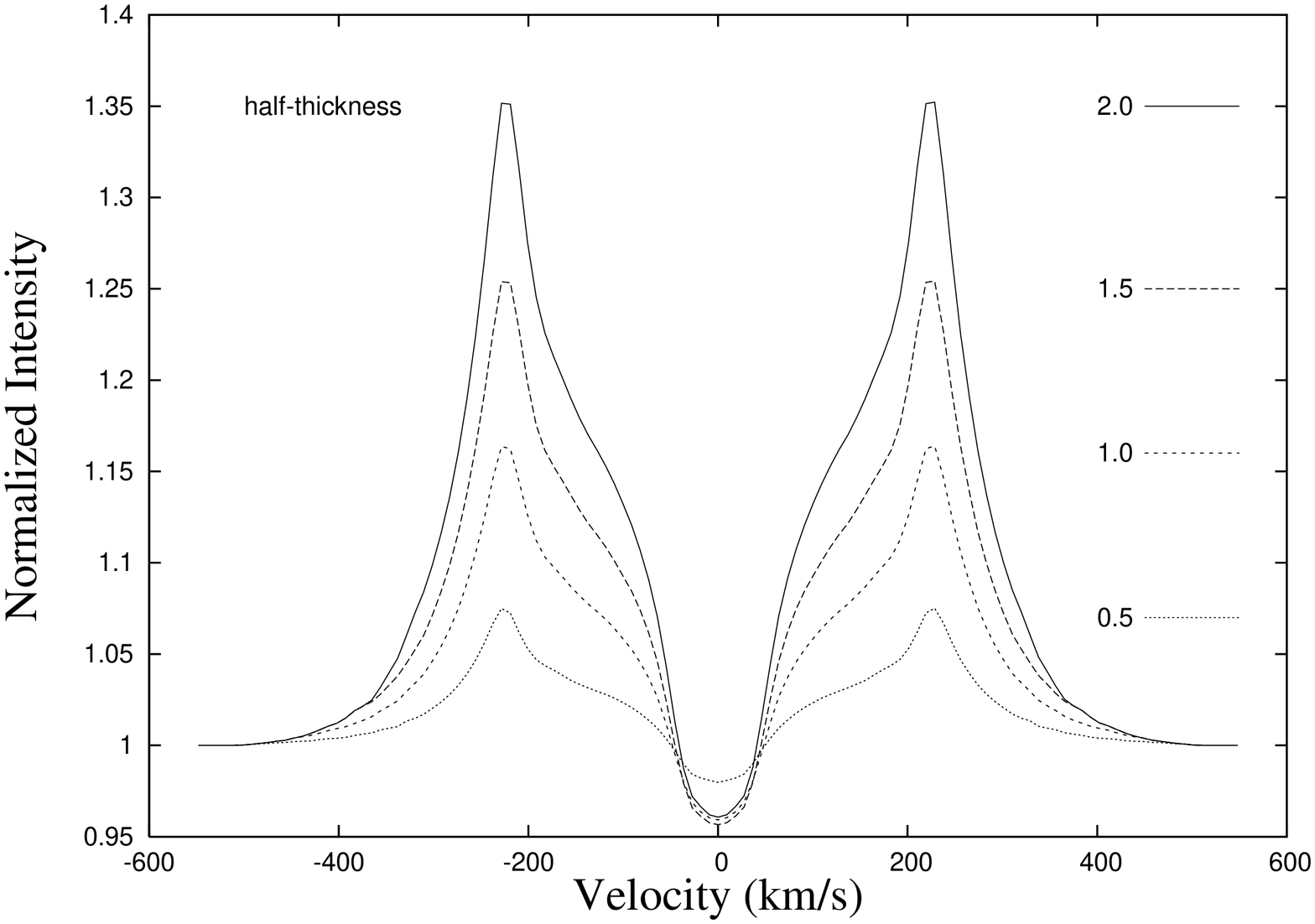}
\includegraphics[width=6.cm,angle=-90,clip=]{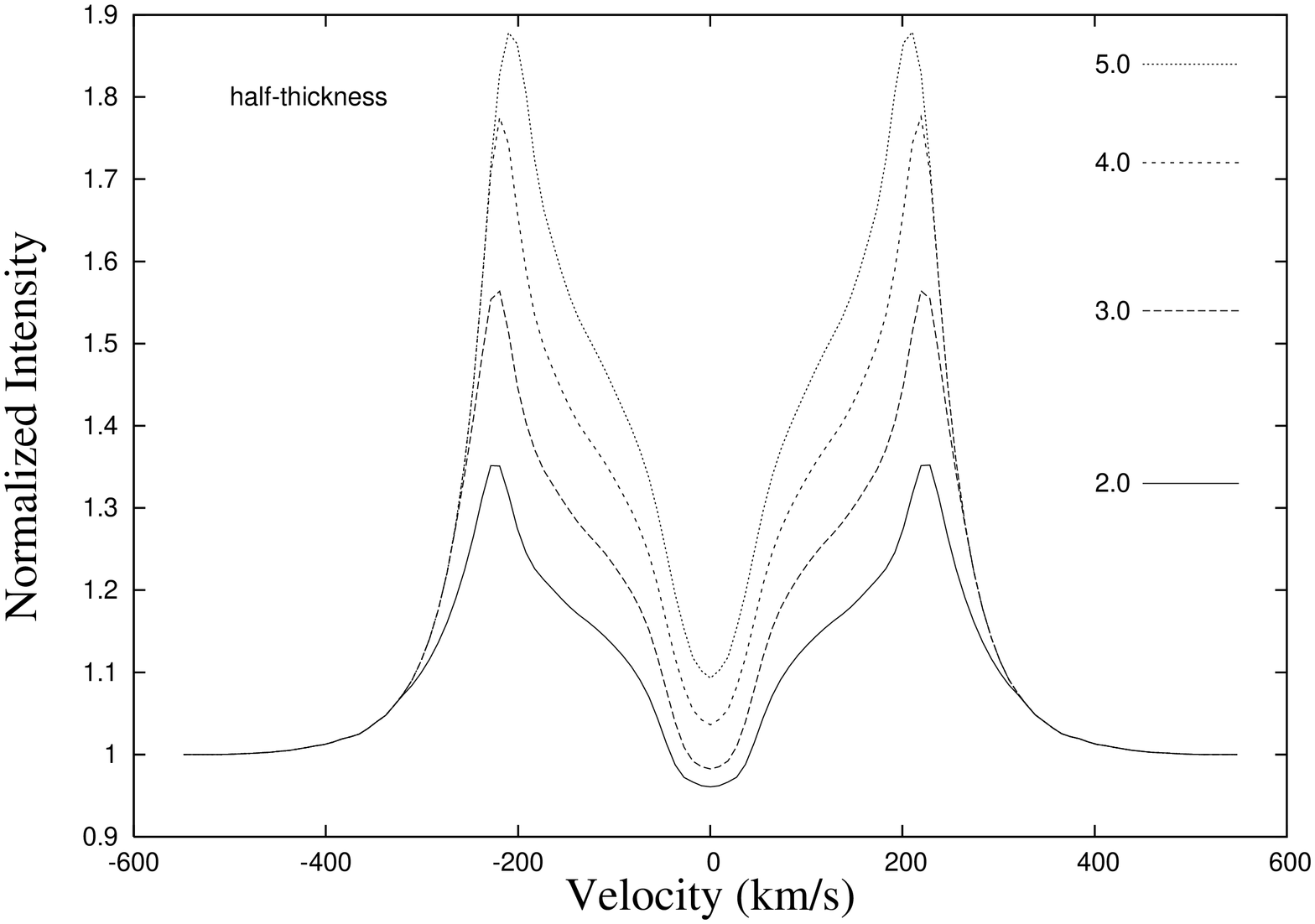}}
\centerline{
\includegraphics[width=6.cm,angle=-90,clip=]{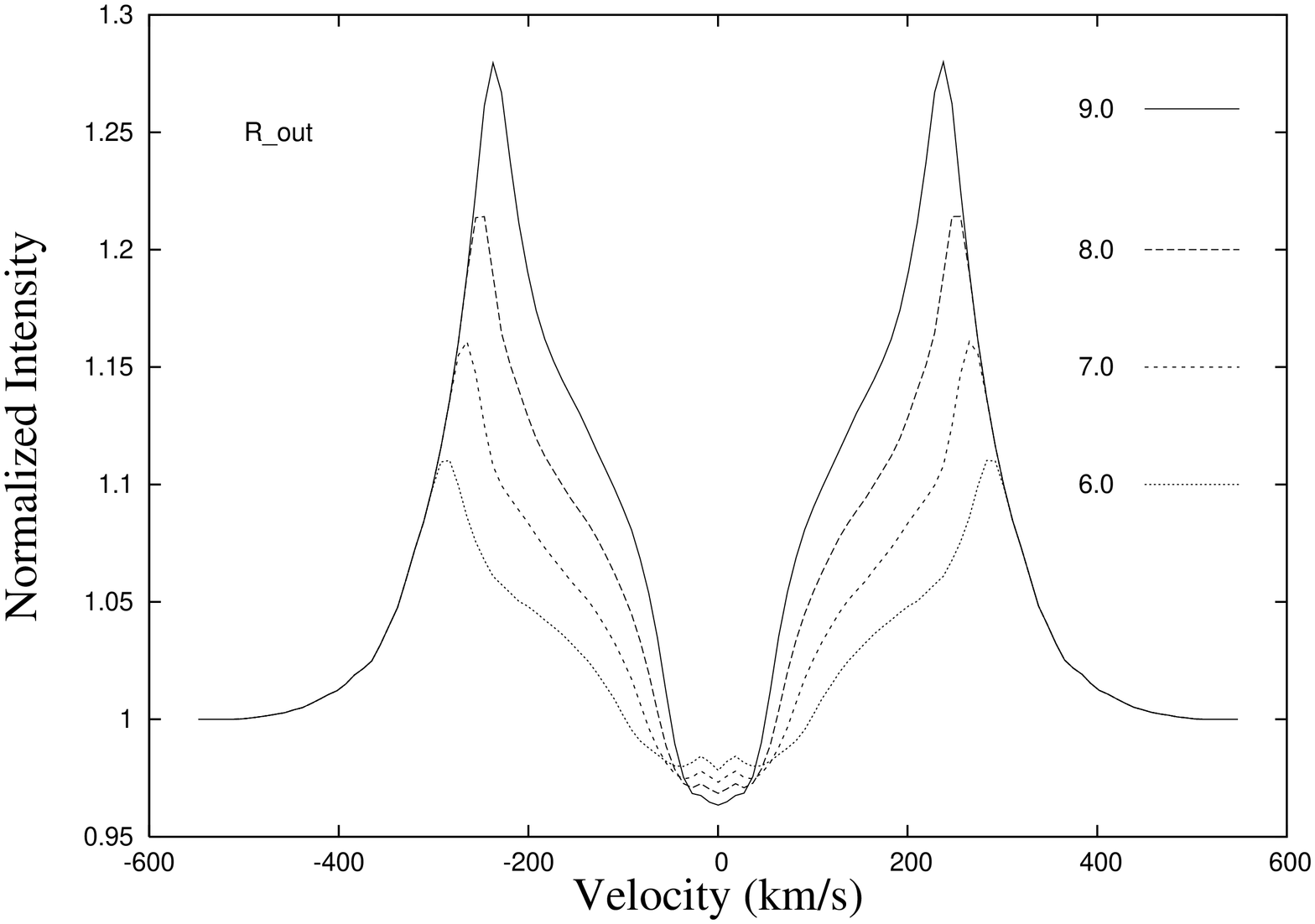}
\includegraphics[width=6.cm,angle=-90,clip=]{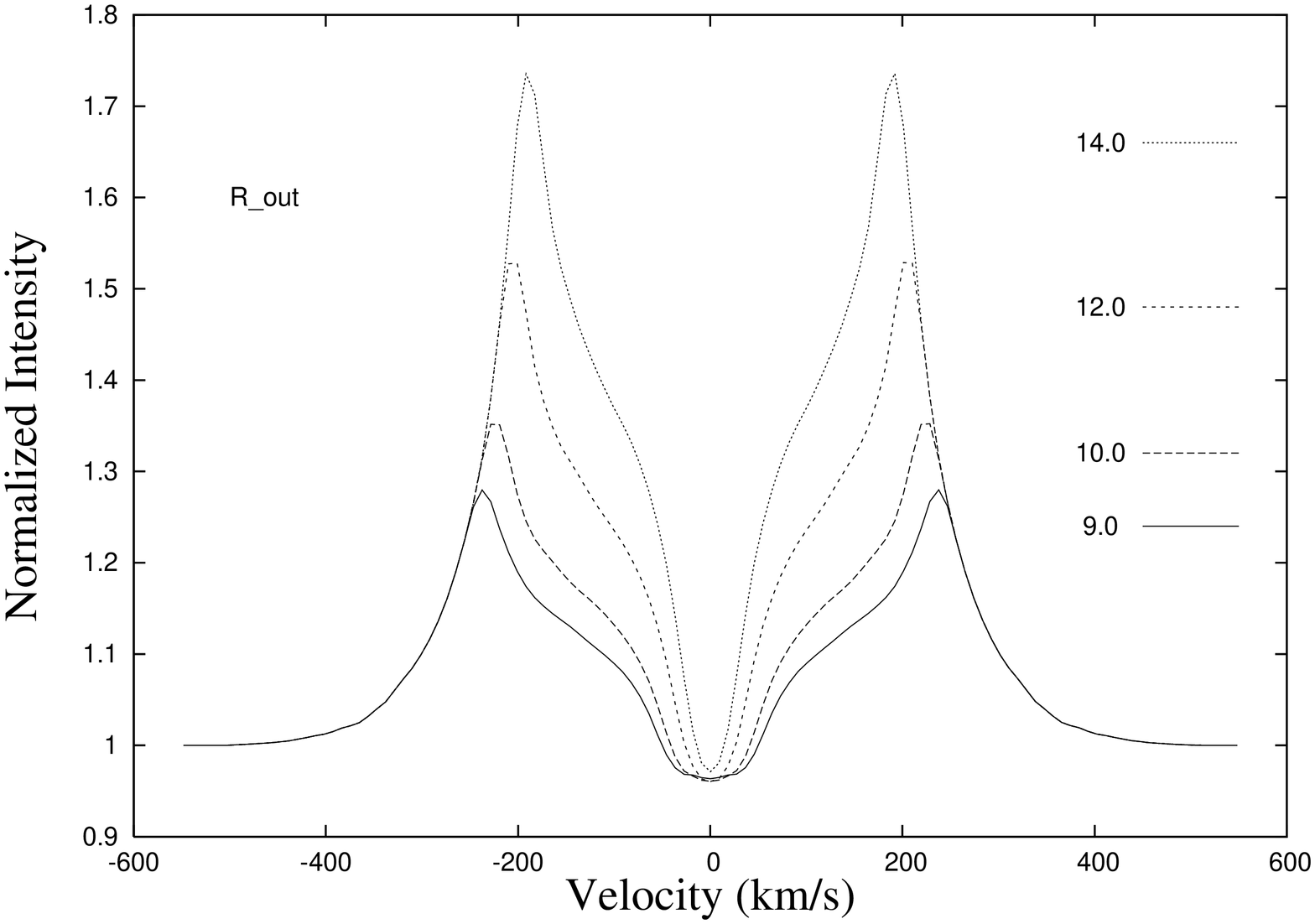}}
\centerline{
\includegraphics[width=6.cm,angle=-90,clip=]{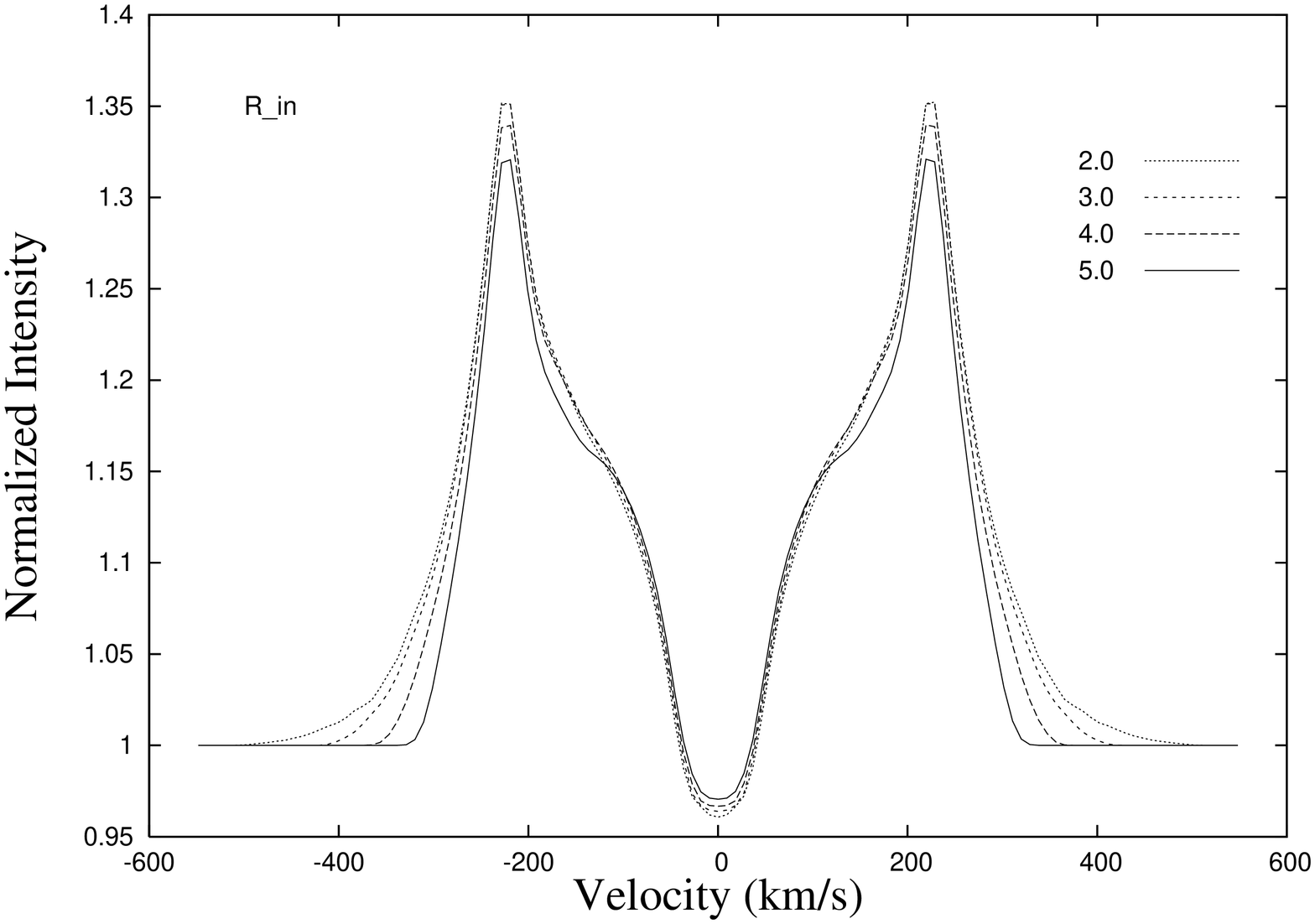}
\includegraphics[width=6.cm,angle=-90,clip=]{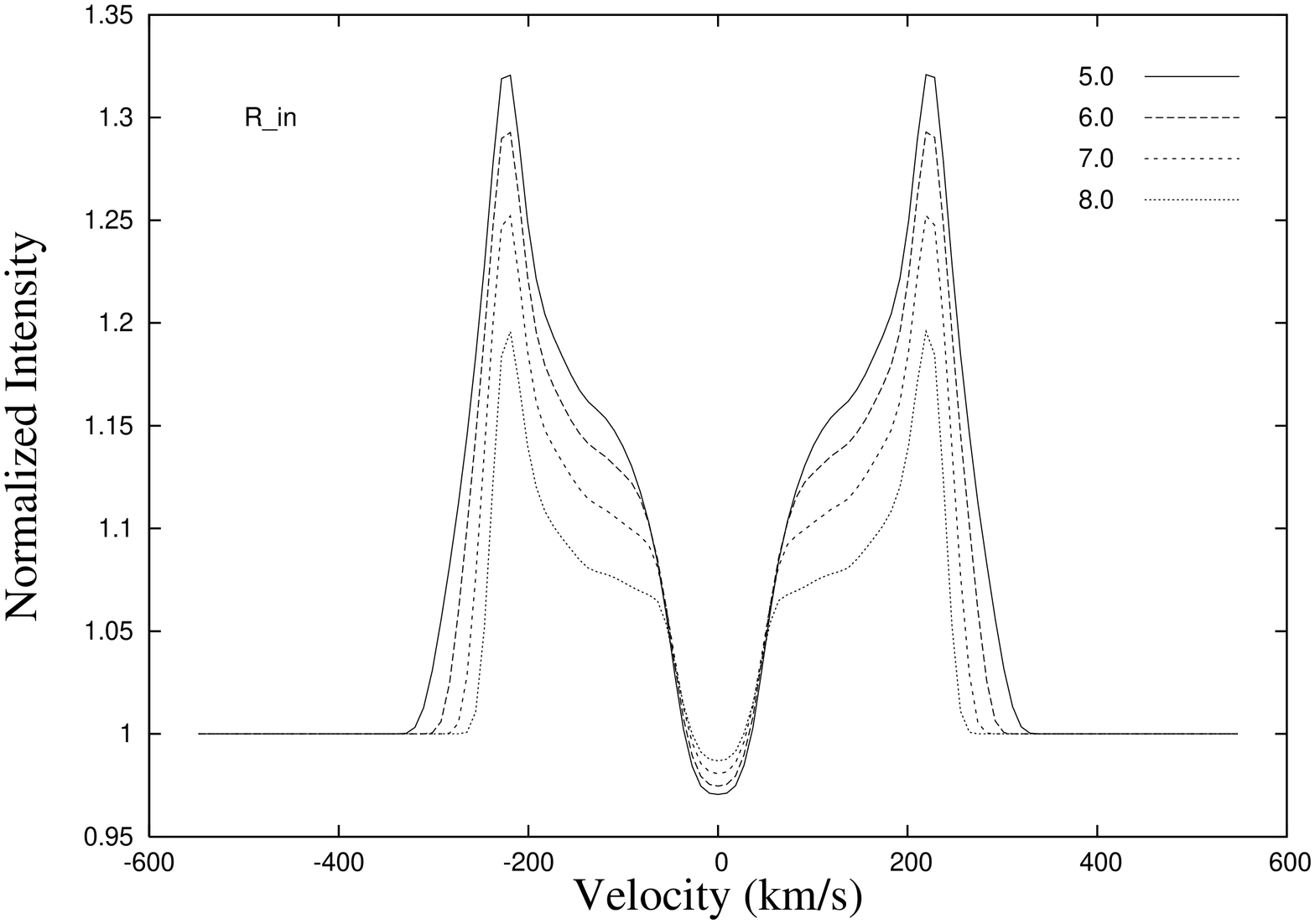}}
\caption{
Top-left: Effect of varying alpha (vertical half-width of the disc)  
from 0.5 to 2 $R_{\odot}$; 
top-right: Effect of varying vertical half-width of the disc 
from 2 to 5 $R_{\odot}$;
middle-left: Effect of varying the outer radius of the disc from 6 to 9
$R_{\odot}$;
middle-right: Effect of varying the outer radius of the disc from 9 to 14
$R_{\odot}$;
bottom-left: Effect of varying the inner radius of the disc from 2 to 5
$R_{\odot}$;
bottom-right: Effect of varying the inner radius of the disc from 5 to 8
$R_{\odot}$.
}
\label{f1}
\end{figure*}

\begin{figure*}
\centerline{
\includegraphics[width=6.cm,angle=-90,clip=]{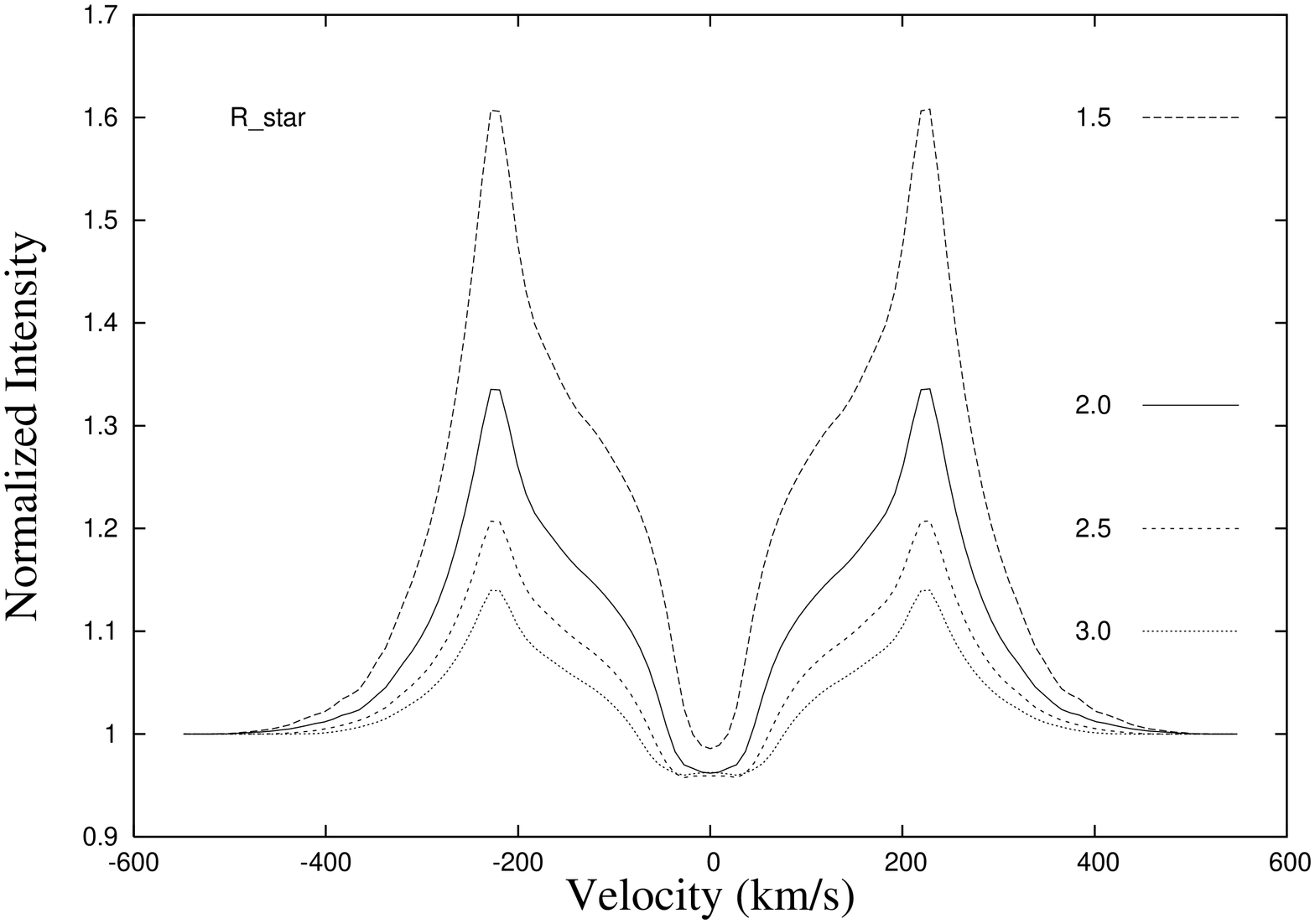}
\includegraphics[width=6.cm,angle=-90,clip=]{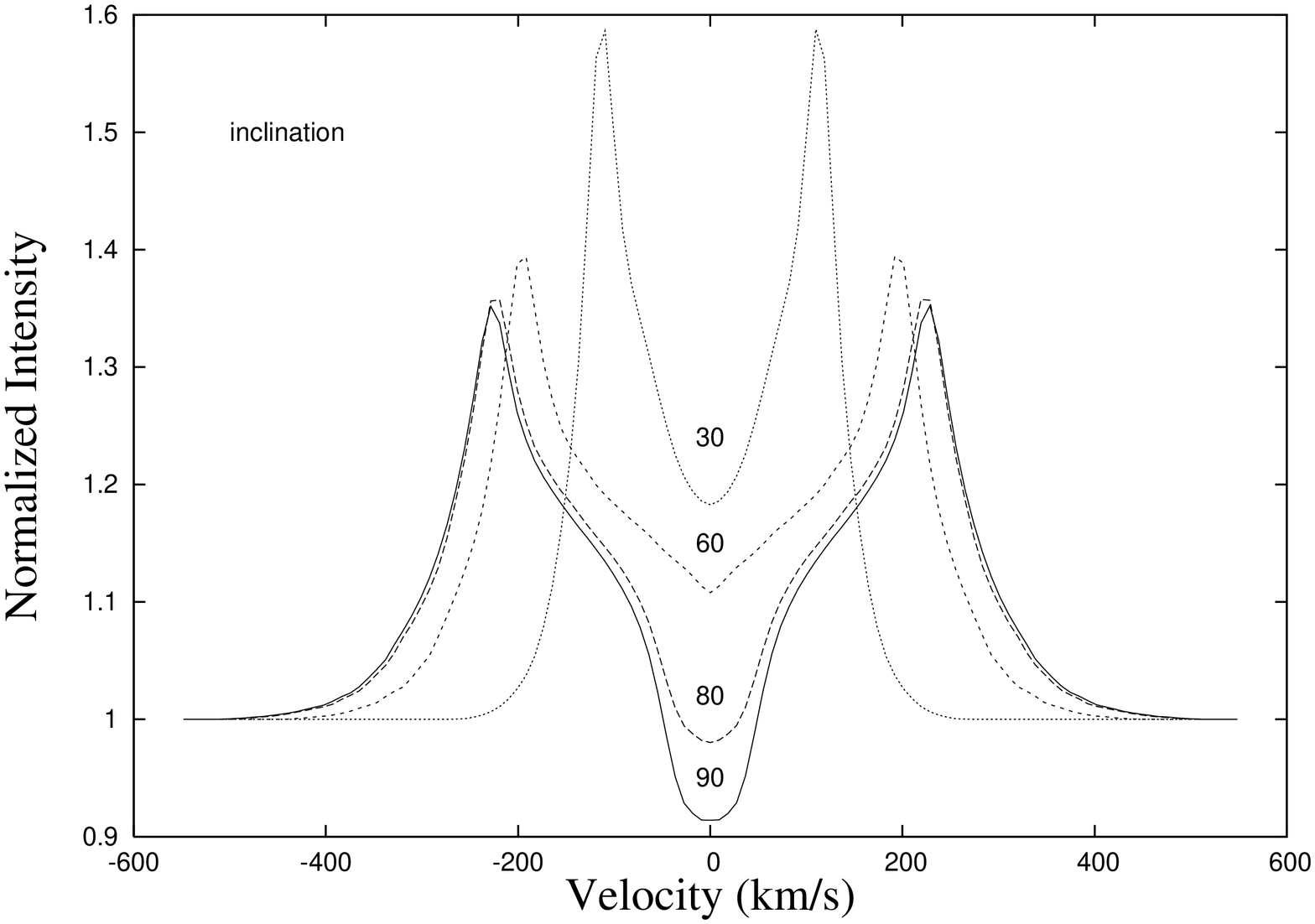}}
\centerline{
\includegraphics[width=6.cm,angle=-90,clip=]{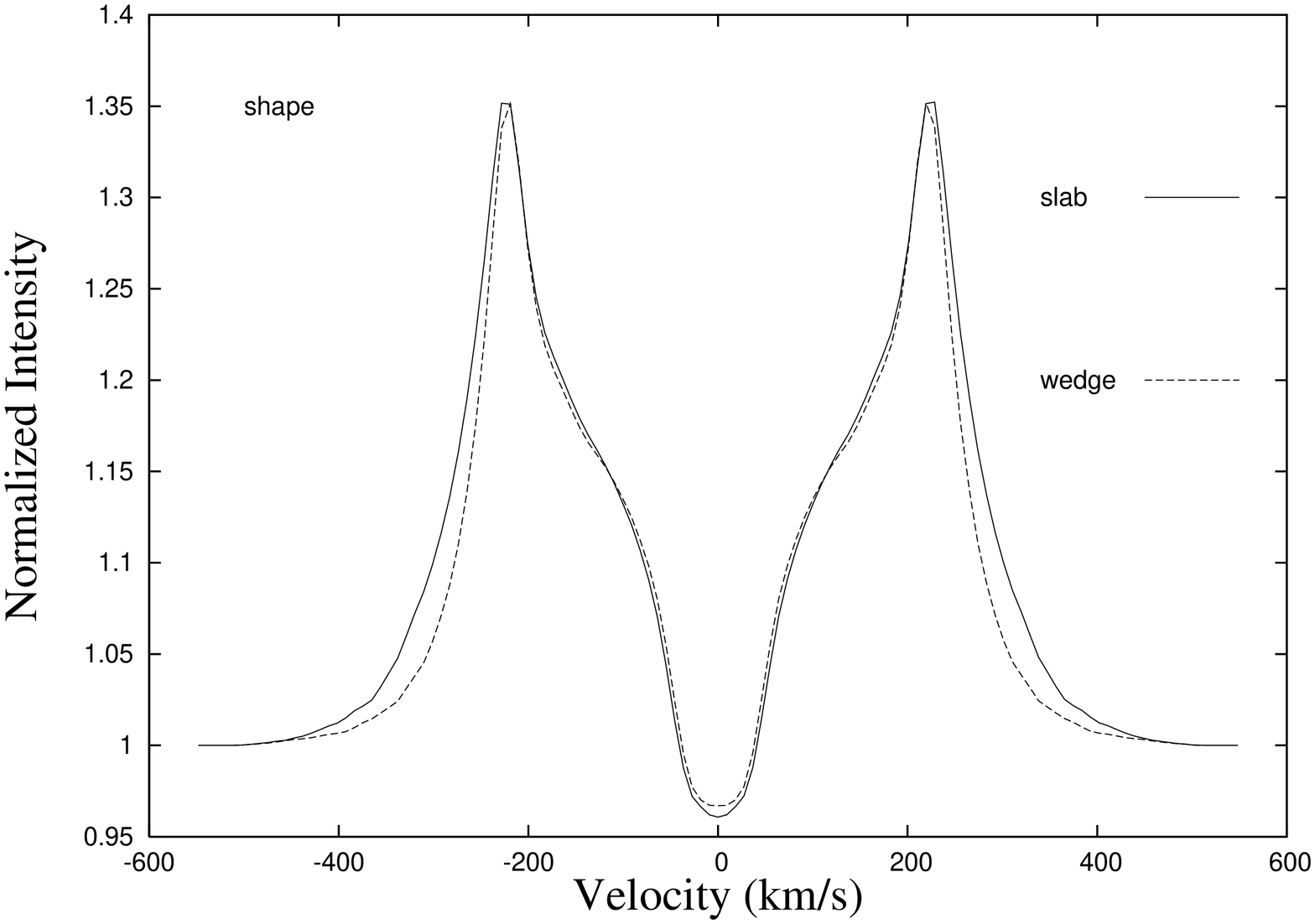}
\includegraphics[width=6.cm,angle=-90,clip=]{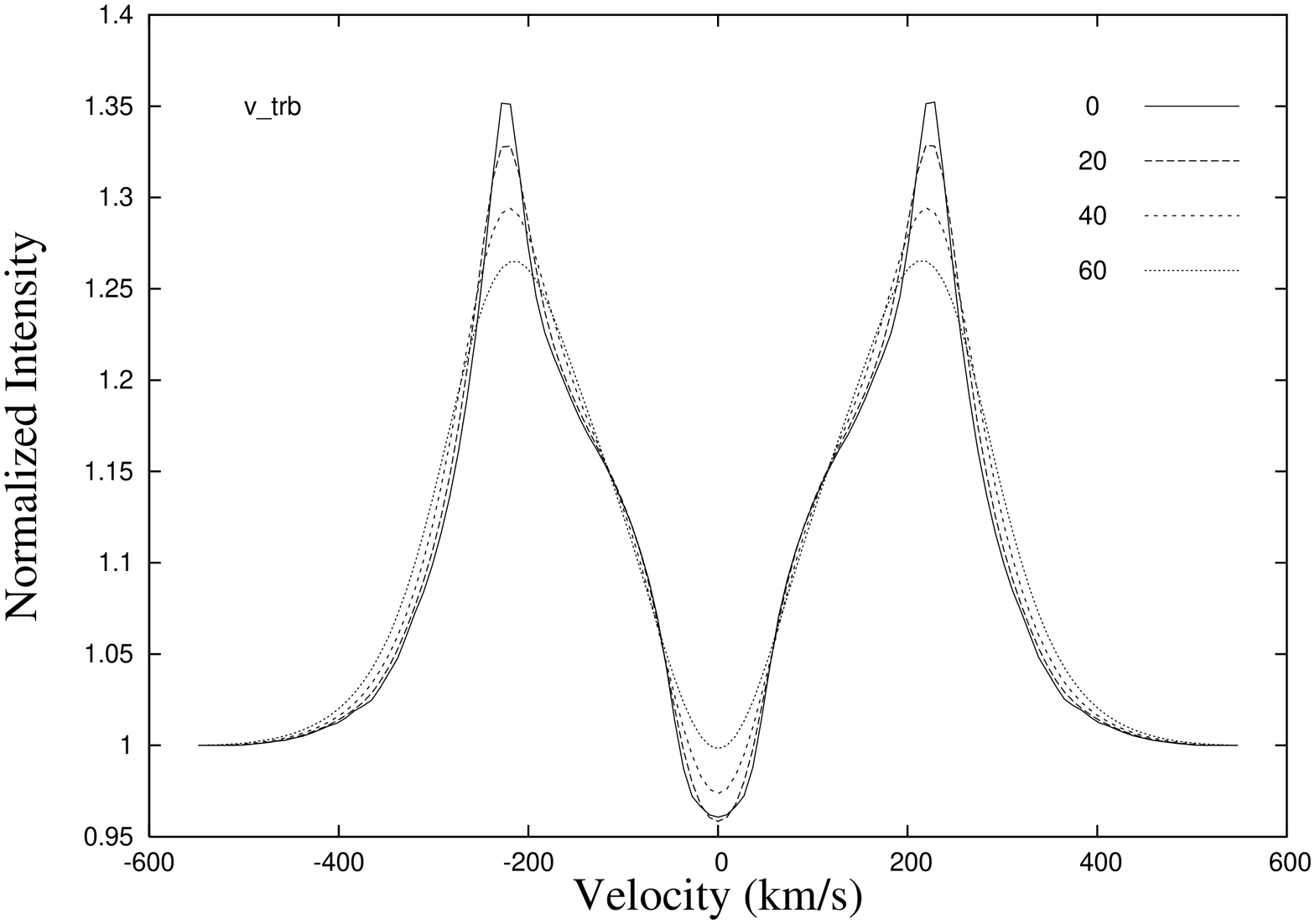}}
\caption{
Top-left: Effect of varying the radius of the star 
from 1.5 to 3.0 $R_{\odot}$;
top-right: Effect of changing the inclination of the disc
from $90^{\circ}$ to $30^{\circ}$;
bottom-left: Effect of the geometrical shape of the disc;
bottom-right: Effect of increasing the microturbulence in the disc
from 0 to 60 km\,s$^{-1}$.
}
\label{f2}
\end{figure*}

\begin{figure*}
\centerline{
\includegraphics[width=6.cm,angle=-90,clip=]{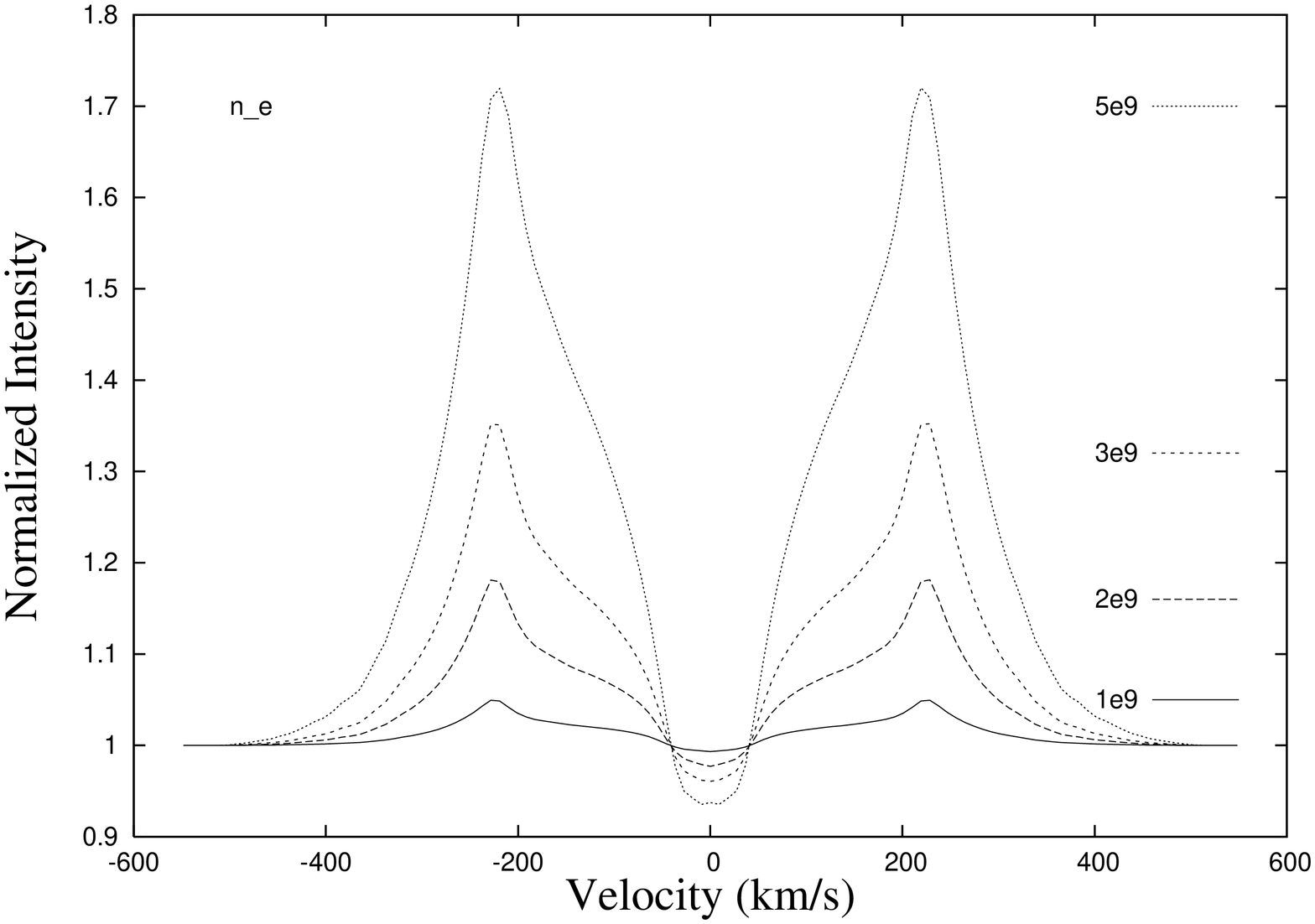}
\includegraphics[width=6.cm,angle=-90,clip=]{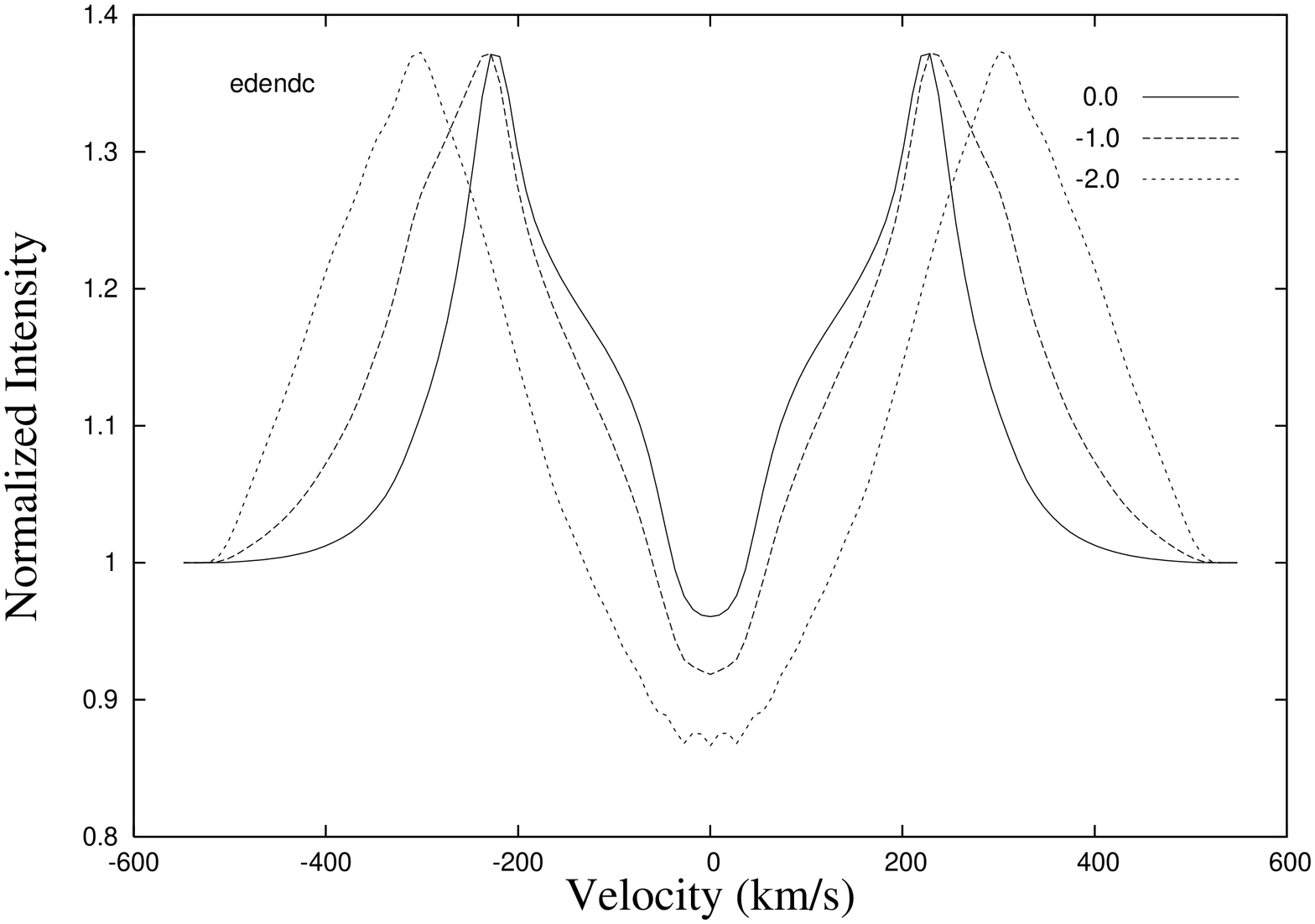}}
\centerline{
\includegraphics[width=6.cm,angle=-90,clip=]{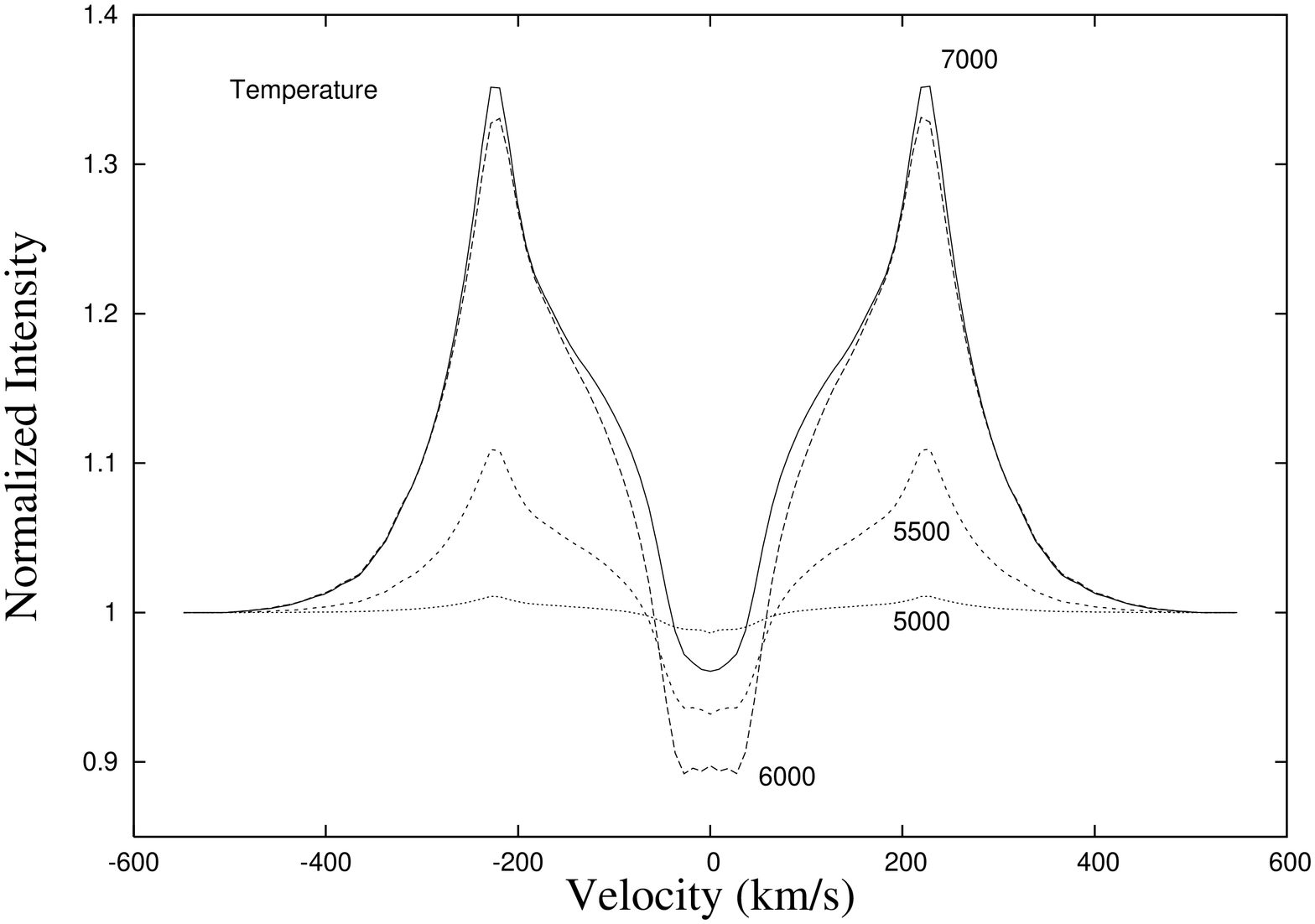}
\includegraphics[width=6.cm,angle=-90,clip=]{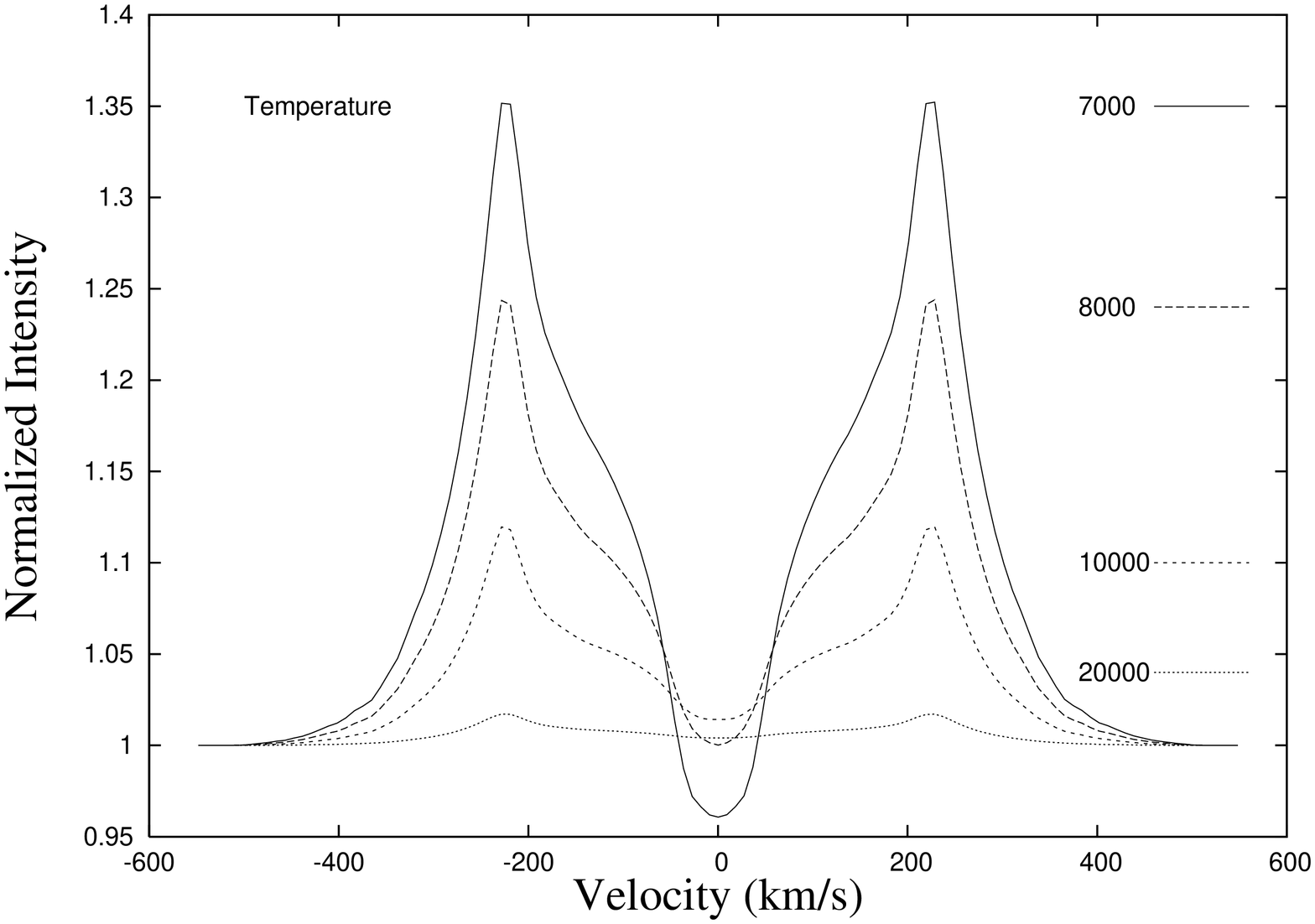}}
\caption{
Top-left: Effect of varying the density and the electron number density
of the disc. $n_{e}$ varies from $1\times 10^{9}$ to 
$5\times 10^{9} cm^{-3}, \rho \sim n_{e}$;
top-right: Effect of varying the exponent $\eta$ of the power law behavior
of the densities $\rho \sim r^{\eta}$ in the disc from 0 to -2;
bottom-left: Effect of varying temperature of the disc from 5000K to 7000K;
bottom-right: Effect of varying temperature of the disc from 7000K to
20000K.
}
\label{f3}
\end{figure*}
%\clearpage

\subsection{The Effects of Various Free Parameters on the Emerging Spectrum}
\label{s2}
Before we modeled the complex spectrum of the two stars and 
the disc, we studied the effects of various free parameters
on the spectrum. To comprehend the problem more easily we restricted 
our calculations to the following assumptions. We neglected
the secondary and studied only the primary star surrounded by a disc.
The primary was treated as a spherical blackbody with the following 
parameters: 
effective temperature $\teff=9800K$,
mass $M=2.63M_{\odot}$, radius $R=1.95R_{\odot}$ (Van Hamme \& Wilson 1993)
and limb darkening $u=0.5$.
The disc was characterized by a slab of
vertical half-width $\alpha=2R_{\odot}$ which was further constrained by 
two spherical surfaces with the inner radius $R_{in}=2R_{\odot}$ and 
the outer radius $R_{out}=10R_{\odot}$.
The electron number density of the disc was set to 
$n_{e}=3\times 10^{9}$ cm$^{-3}$,
density $\rho=5\times 10^{-15}$ g\,cm$^{-3}$, temperature $T=7000$K,
microturbulence $v_{trb}=0$ km\,s$^{-1}$, and inclination $i=82.84^{\circ}$.
We adopted the following velocity field ${\boldsymbol v}$ of the disc:
\begin {equation}
\omega(r)=\sqrt{G\frac{M}{r^{3}}}
~~~~~~~~~~~~~~{\boldsymbol v}=\boldsymbol{\omega} \times {\boldsymbol r}
\label{e1}
\end {equation}
where $M$ is the mass of the central object, 
$G$ is the gravitational constant, ${\boldsymbol \omega}, {\boldsymbol r}$ 
are the angular velocity and radius vectors, respectively.
The above prescription for the velocity reduces to the well known 
Keplerian velocity field in the limit of zero vertical thickness of the disc.
The calculations were performed on the Cartesian 3D grid with 
the typical resolution in the disc plane of $121\times 121$ points.
Figures \ref{f1} to \ref{f3} illustrate the effects of varying just one of 
these free parameters of the system.

The top of Figure \ref{f1} displays the effect of increasing 
the vertical half thickness of the disc from $0.5$ to $5 R_{\odot}$.
The overall emission increases as we increase the emitting space volume,
and the central absorption decreases mainly for the thicker discs.
This occurs because the central depression is partly produced by  
the material projecting onto the star and there is more material in
the Keplerian disc which does not project onto the stellar disc (surface)
for thicker discs, thereby filling the central absorption.
This is not true for very thin discs as it essentially
causes the disc and its line profile to diminish and both 
emission and central absorption approach the continuum level. 
Contrary to the case of the wedge-shaped disc (Budaj \& Richards 2004), 
we observe that in a slab disc geometry 
the blue/red line profile wings of the thick discs ($\alpha>3R_{\odot}$)
do not change much and emission peaks get slightly closer.
 
The middle section of Figure \ref{f1} displays the effect of 
changing the outer radius of the disc from 6 to 14 $R_{\odot}$. 
The emission increases as the emission volume increases.
This parameter has a strong effect on the position
of the emission peaks and width of the central depression. 
The peaks get closer and the central depression narrower for larger
radius as more matter is involved at lower Keplerian velocities.
The depth of the central absorption varies very little since the amount of
the matter projected onto the stellar surface does not vary as much as
in the previous case. This absorption first deepens as the overall 
emission becomes more pronounced and then for $R_{out}>12R_{\odot}$ 
weakens as the fraction of the gas projecting on the stellar disc
decreases. The blue and red wings of the line are not affected at all.

The bottom part of Figure \ref{f1} displays the effect of increasing 
the inner radius of the disc from 2 to 8 $R_{\odot}$. 
Contrary to the outer radius, the inner radius
has little effect on the emission from the beginning when it is a small
fraction of the outer radius since it does not change the emission volume
noticeably.
The changes are mainly seen in the far wings of the profile since
the higher Keplerian velocities are involved.
Only when it approaches the outer radius, the overall emission starts
to decline and the line weakens as the emission volume decreases, and
the shape of the profile acquires the new U-type shape with a central hole.
This shape indicates that the central depression is created by two
different mechanisms (see below).

In Figure \ref{f2} (top left frame), we inspect what happens 
if we vary the stellar radius from 1.5 to 3.0 $R_{\odot}$.
Stellar radius certainly changes the amount of matter projected onto 
the stellar surface but it also raises the continuum level and dilutes 
the emission from the disc. This is the main reason why the emission
weakens for bigger stars. At the same time, 
the central absorption gets wider because
the matter projected onto the stellar disc spans a broader radial 
velocity interval. When varying the stellar radius, it may happen
that the star overlaps the inner radius of the disc. 
In such a case, the program sets both radii to be equal.

We also illustrate here the line profiles if the disc were viewed from 
different inclination angles (Figure \ref{f2}, top right frame).
Starting from edge-on, $i=90^{\circ}$, the two emission peaks have
the lowest intensity and we also observe the largest 
separation of the emission peaks and the deepest central depression. 
This is because we encounter the highest radial velocities and 
most of the disc which is cooler than the star projects onto 
the stellar surface.   
As the inclination decreases, and we begin to see the disc pole-on,  
the emission peaks increase and their separation and central depression
vanish, eventually merging into single peak emission for $i=0^{\circ}$. 
One can also see that the central depression is caused by two different
effects.
The velocity field and geometry of the disc itself can produce
the U or V type of the depression.
Superposed on that is the absorption by the cool matter projected
onto the hotter stellar surface which quickly diminishes if viewed 
out of the disc plane.

It is interesting to explore how the shape of the profile changes
if the geometry of the disc changes. The prescription for the velocity
field remains the same. In Figure \ref{f2} (bottom left frame), we compare 
two profiles: one obtained for the slab shaped disc with 
the standard parameters listed above. The other disc has the shape of a
rotating wedge with the wedge angle $\alpha=28^{\circ}$, which produces
emissions of similar strength to the slab disc model.
All other parameters were the same. It is clear that the main
difference is in the wings, which are more pronounced in the slab
model since there is more matter closer to the star in that model
than in the wedge model. 

Apart from the state quantities and velocity field, every space point can
be assigned a value of microturbulence. It is included in 
the calculation of thermal broadening as an additional thermal motion
and can thus be used to model chaotic velocity fields on the scale   
smaller than the mean free path of a photon.
It turns to be a useful free parameter since the mass transfer may not
be a smooth process and the velocity field of the spiraling gas
in the disc may well depart from the circular Keplerian orbits and be
turbulent, and these departures may easily exceed the sound velocity  
which is of the order of 10 km\,s$^{-1}$.
Figure \ref{f2} (bottom right frame) illustrates the effect of increasing 
the microturbulence in the disc from 0 to 60 km\,s$^{-1}$.
Note that the inner radius Keplerian disc velocity is about 500
km\,s$^{-1}$.
The emission peaks are gradually smoothed and are broader.
The behavior of the central absorption is more complicated and interesting.
From the beginning it gets narrower and even deeper because of 
the desaturation effects of the microturbulence but then the smoothing 
effect prevails.
The microturbulence affects the optical depth along the line of sight.
While the surface of the star is seen at an optical depth of approximately
1.5 (at the frequency of the line center) without the microturbulence,
it is seen at the optical depth of 0.7, 0.4 for the microturbulence of  
20 and 40 km\,s$^{-1}$, respectively. 
Without the turbulence and assuming the input parameters mentioned above,
the line opacity was found about 3 orders of magnitude higher than 
the continuum opacity in the line center. 
Thomson scattering was the most important continuum opacity source and
was a factor of 300 higher than the hydrogen bound-free contribution, 
while H bound-free opacity was a factor of 20 higher than the hydrogen
free-free opacity, and H free-free opacity a factor of 30 higher than 
Rayleigh scattering opacity in the disc. This pattern, however, will be
very sensitive to the state quantities, wavelength, and line absorption 
coefficient.

Figure \ref{f3} (top left frame) displays the effect of increasing 
the electron number density of the disc from $1\times10^{9}$ to 
$5\times10^{9}$cm$^{-3}$.   
Here, we assume that hydrogen is almost fully ionized,
in which case the density is linearly proportional to the electron number
density. This strongly enhances the emission peaks and slightly
deepens the central absorption. This huge impact on the emission
can be understood since the equivalent width in the optically thin case
is proportional to the population of the particular level, which in turn
is proportional to the total HI population. However, the ionization 
fraction of neutral hydrogen is proportional to the electron number 
density, and if the latter is also proportional to the density it   
follows that the total HI population increases with the square of   
the density or the electron number density assuming the hydrogen    
abundance and temperature are fixed 
($n_{HI}/n_{HII}\approx n_{HI}/n_{H}\sim n_{HI}/\rho \sim n_{e}\sim \rho$).
As a consequence,
the equivalent width, EQW, should behave also like
$\sim n_{HI} \sim \rho^{2}\sim n_{e}^{2}$ 
which is seen in Figure \ref{f3}.
 
Since the emission is so sensitive to the density we also studied 
the effect of different density profiles (Figure \ref{f3}, top right frame). 
Let the density and electron number density be inhomogeneous
and have a power law dependence on the distance from the star,
namely, $\rho\sim r^{\eta}$. This was achieved by varying the exponent
$edendc\equiv \eta$ from 0.0 to -2.0 and by normalizing the synthetic
spectrum to the same emission peak strength.
It is apparent from the figure that the line profiles for lower exponents
are broader and have steeper wings while those for higher exponents have 
narrower and smaller central depression. It can easily be explained 
since the lower exponents emphasize    
the matter close to the star with higher Keplerian velocity while the      
opposite is true for the higher exponents.
Contrary, to what was found by Budaj \& Richards (2004) for a wedge-shaped
disc with the inner radius of 2.5$R_{\odot}$ the position of the emission 
peaks in a slab disc model with $R_{in}=2.R_{\odot}$ is much
more sensitive to this exponent.

Figure \ref{f3} (bottom left and bottom right frames) displays also 
the effect of varying the temperature of the disc over the interval  
5000-20000K. 
The emission is strongest at about 7000K and declines
towards higher temperatures as the fraction of neutral hydrogen declines
due to ionization. The emission also declines towards cooler temperatures
as it decreases the population of the lower level from which 
the $H\alpha$ originates.
At the same time, the temperature of the disc has a strong effect on
the depth of the central absorption which grows towards
the cooler temperatures.  For temperatures $T<\teff$, 
the line source function along the line of sight 
hitting the stellar surface steps down from the stellar surface to disc 
and towards the observer. For higher temperatures $T>\teff$, 
this component of the source function steps up reducing the central 
depression.
For both cooler and hotter temperatures than 7000K, 
the line-to-continuum emissivity decreases 
(the continuum is provided by the star mainly and is essentially constant)  
and the line disappears from the spectrum forcing the central depression
to merge with continuum as well. This is the reason why the central 
depression ceases also below 6000K although the line source function 
steps down. The cooler discs if they were more dense could produce even 
deeper central absorption. Nevertheless, one can see that discs or regions 
with temperatures below 5000K or above 30000K are not capable of producing
strong Balmer line emissions and that most of such emissions observed
in the Algols or CVs must originate from the above mentioned temperature 
regions.

We would like to emphasize that these calculations do not aim
to calculate the physical model of the disc. The model is assumed
to be known. The previous calculations illustrate the effect of varying
just one free parameter at a time while keeping all the others fixed.
In many cases this may not represent the real behavior of the system 
since some quantities may be more or less interlocked such as temperatures 
and electron number densities.

\subsection{Modeling of the Observed Spectra of TT Hya} 
\label{s3}
Now that we understand the influence of various parameters on 
the emerging spectrum, we can proceed to fit the complex spectrum of TT Hya. 
We used 23 observed spectra from 1994 of Richards \& Albright (1999)
complemented by 24 new spectra from 1996/97. A detailed study
of these spectra will be published in Richards et al. (2005)
and will include the actual measurements of the intensity of 
the central depression and the emission peaks.

\begin{table}
\begin{center}
\caption{Adopted properties of TT Hya.}
\begin{tabular}{lll}
\hline
\hline
~~~~~Primary:&                &Source\\
$M$	  & $2.63 M_{\odot}$  & VW\\
$R$	  & $1.95  R_{\odot}$ & VW\\
$\log g$  & 4.23              &  \\
$T$	  & 9800 K            & VW,E\\
$v\sin i$ & 168 km\,s$^{-1}$  & E\\
u         & $0.5$             & VW\\
a         & $22.63 R_{\odot}$ & VW\\
\hline
~~~~~Secondary:&              &\\
$M$	  & $0.59 M_{\odot}$  & VW\\
$\log g$  & 2.66              & \\
$T_{p}$	  & 4600 K            & BRM\\
u         & $0.8$             & VW\\
$\beta$   & 0.08              & L\\
\hline
~~~~~Disc:&                   &\\
$i$		& $82.84^{\circ}$     & VW\\
$\alpha$	& $2 R_{\odot}$       & BRM\\
$R_{in}$	& $2 R_{\odot}$       & BRM\\
$R_{out}$	& $10. R_{\odot}$     & BRM\\
$\rho(R_{in})$  & $4\times 10^{-14}$ g\,cm$^{-3}$ & BRM  \\
$n_{e}(R_{in})$ & $2\times 10^{10}$ cm$^{-3}$     & BRM \\
$T$             & 6200 K              & BRM\\
$\eta$	        & $-1$                & BRM\\
$v_{trb}$		& 30 km\,s$^{-1}$     & BRM\\
\hline
\end{tabular}
\label{t2}
\end{center}
Designation: 
$u$ - limb darkening coefficient, 
$a$ - separation of the components,
$T_{p}$ - temperature at the rotation pole,
$\beta$ - gravity darkening exponent,
$i$ - inclination,
$\alpha$ - vertical half thickness, $R_{in}$ - inner radius, 
$R_{out}$ - outer radius, $\rho(R_{in})$ - density, 
$n_{e}(R_{in})$ - electron number density, 
$\eta$ - exponent of the density power law,
$v_{trb}$ - microturbulence. Secondary was assumed to have synchronous
rotation and to fill the Roche lobe,
VW - Van Hamme \& Wilson (1993), 
E - Etzel (1988), 
L - Lucy (1968), 
BRM - this work.
\end{table}

%\clearpage
\begin{figure}
\centerline{
\includegraphics[width=19.cm,height=9.cm,angle=-90,clip=]{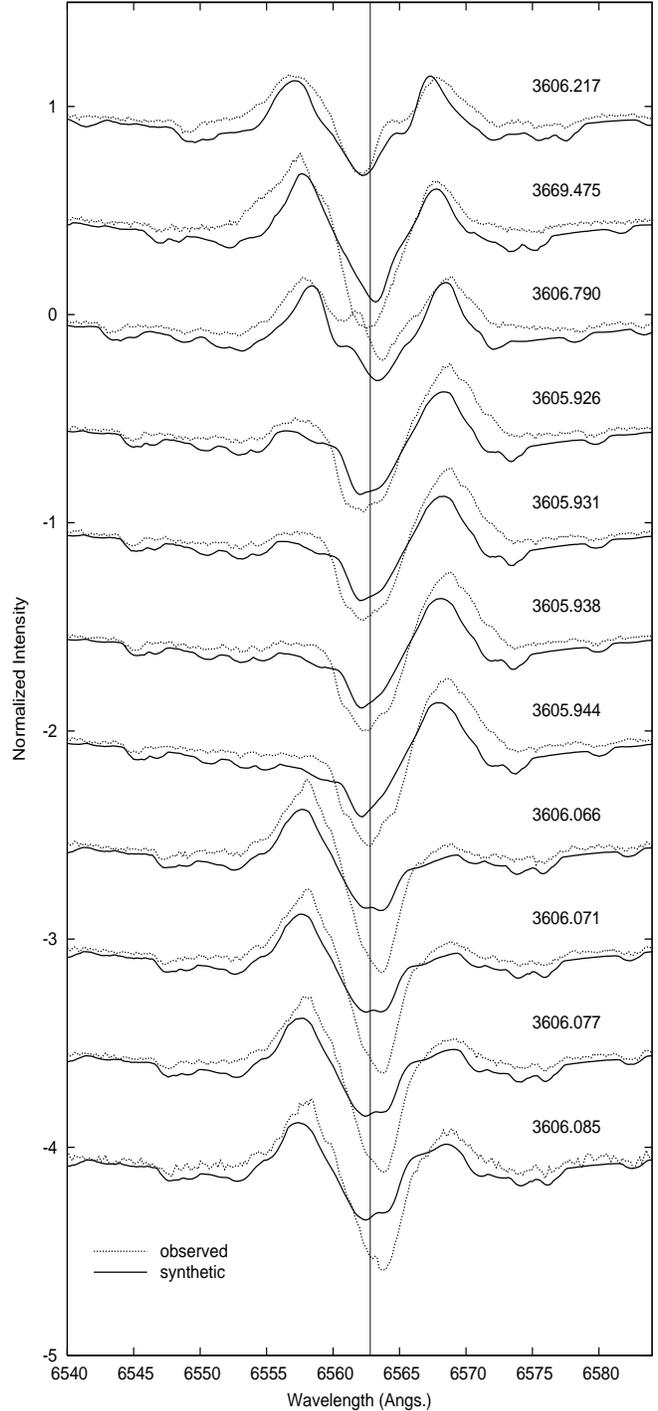}}
\caption{Synthetic spectra (solid line) compared with the observed spectra 
(dotted line) of TT Hya at different phases. The vertical line indicates
the laboratory wavelength. The systemic (gamma) velocity of the system 
(9.25 km\,s$^{-1}$) was subtracted from the observed spectra. 
The numbers to the right of the spectra indicate the epoch and phase
of the observations.}
\label{f4}
\end{figure}

\begin{figure}
\centerline{
\includegraphics[width=15.cm,height=9.cm,angle=-90,clip=]{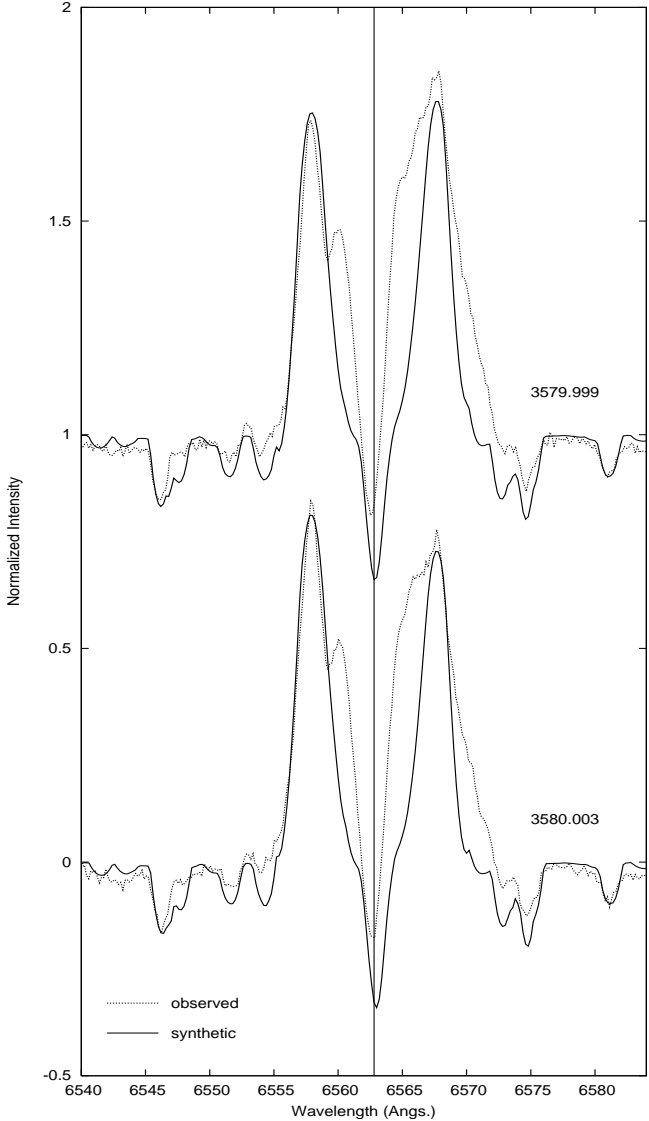}}
\caption{Synthetic (solid line) and observed spectra (dotted) of TT Hya 
during total eclipse. The numbers to the right of the spectra indicate the
epoch and phase of the observations. The vertical line indicates 
the laboratory wavelength. The systemic (gamma) velocity of the system 
(9.25 km\,s$^{-1}$) was subtracted from the observed spectra.
}
\label{f5}
\end{figure}
%\clearpage

We performed some new and more sophisticated calculations of synthetic
spectra in which both stars and a disc were considered.
The primary was treated as a rotating solid sphere while the surface
of the secondary was modeled in terms of the Roche model.
The stellar properties were based on the absolute
parameters of the system obtained by Van Hamme \& Wilson (1993)
and are summarized in Table \ref{t2}. Since our code accepts
the polar temperature, $T_{p}$, for the Roche model rather than 
the average surface temperature, we adopted the value $T_{p}=4600$K 
which is slightly higher than the average temperature $T=4361$K 
obtained by Van Hamme \& Wilson.
Their V-band limb darkening coefficients were adopted here.
The surface gravity, $\log g=4.23$ for the primary and
$\log g=2.66$ for the secondary were assumed.
First, we obtained the model atmospheres of both stars by interpolating 
in the Kurucz (1993b) $T_{\rm eff}-\log g$ grid and assuming solar 
abundances and a microturbulence of 2 km\,s$^{-1}$. Next, we calculated 
the intrinsic synthetic spectra emerging from these atmospheric models 
(flux from the unit area of the surface) using the code {\sc{synspec}} 
(Hubeny et al. 1994) modified by Krti\v{c}ka (1998).
Here again, solar abundances and a microturbulence of 2 km\,s$^{-1}$ 
were assumed.
These synthetic intrinsic spectra were then assigned to the primary and
secondary for the calculation of the complex spectra of both stars and 
the disc using {\sc{shellspec}}. Based on the lessons learned in 
the previous section 
we manipulated the free parameters of the disc so that the observed spectra 
could be reproduced. We assumed that the electron number density is
equal to the hydrogen number density. A good fit to the observations was 
achieved with the realistic parameters for the disc. These values are 
also summarized in Table \ref{t2}. 

Figures \ref{f4} and \ref{f5} depict both the observed spectra 
as well as our synthetic fits for a few representative phases at both
quadratures and during both eclipses to illustrate the most interesting 
effects. The observed spectra display red and blue emission features
as well as a central depression. 
The observed far wings of $H\alpha$ stretching beyond 
the emission peaks in Figure \ref{f4} do not seem as deep as in 
the synthetic spectra. This is not a continuum rectification problem.
These shallower observed wings were previously noted by Peters (1989) 
and interpreted as an emission stretching up to 30\AA\ from the line center.
However, these observed far $H\alpha$ wings in TT Hya differ from 
single star standard spectra or from synthetic ones most probably 
because the atmospheric regions of the primary are more complicated.
A number of additional effects could be expected here like
a suction (``spray gun'') effect of a disc moving over the atmosphere, 
centrifugal forces, and disc-atmosphere interaction with higher
turbulence or temperatures.
All of them tend to reduce the effective gravity of the atmosphere
of the primary and the atmosphere-disc transition region.  
Note, that the effective gravity is essentially zero in the disc.
This could change $T-\tau$ behavior of the atmosphere, mimic lower 
$\log g$, reduce the pressure broadening, and, consequently, cause 
shallower line wings or shell spectra. 
It might be misleading to interpret these shallower wings as an emission
(Peters 1989). 
The emission peaks and the central depression depend strongly on the phase, 
as best illustrated in Figure \ref{f6}, for all observed and synthetic 
spectra. A discussion of the phase dependence of $H{\alpha}$, 
simultaneously on Figures \ref{f4}, \ref{f5}, and \ref{f6} follows.

\subsubsection{First Quadrature}
The first quadrature (Figure \ref{f4}: phase 0.217) 
seems to have the simplest spectrum. Here, we see a remarkably good 
agreement between the observed and theoretical profiles. 
Both emission peaks have approximately similar strength. 
A little bump on the red side of the central depression is the intrinsic 
$H\alpha$ line of the secondary.

\subsubsection{Secondary Eclipse}
\begin{figure*}
\centerline{
\includegraphics[width=12.cm,height=18.cm,angle=-90,clip=]{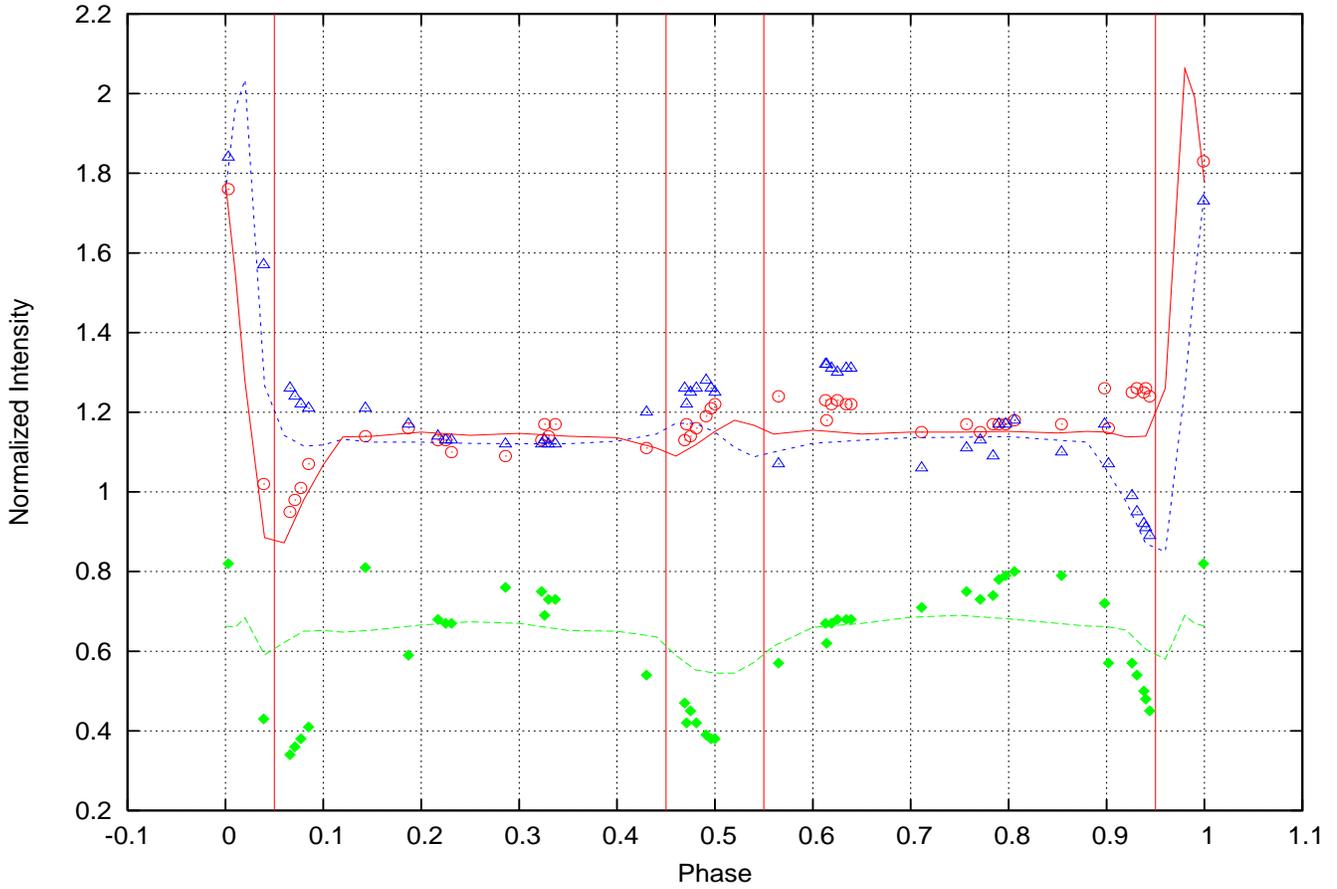}}
\caption{Behavior of the emission strength and central absorption
depth during the orbit of TT Hya.
Blue triangles -- highest normalized intensity of the blue emission peak;
red circles -- intensity of the red emission;
green diamonds -- depth of the central depression;
green long-dashed line -- synthetic central depression;
solid red line -- synthetic red emission;
blue short dashes -- synthetic blue emission.
The vertical lines indicate the start and the end of primary and secondary 
(partial) eclipse.}
\label{f6}
\end{figure*}

As we proceed towards secondary eclipse,
the blue emission gets apparently stronger than the red emission.
This should not happen if the disc is symmetric and its source function
does not depend on the phase. However, this effect is observed also in 
the synthetic spectra (see Figure \ref{f6} and Figure \ref{f4}: phase 0.475) 
and is caused by the shadowing of the secondary by the disc. 
Just before secondary eclipse the receding parts (red emission) 
of the disc projects onto the surface of the secondary. This imposes 
different boundary conditions for intensity which is zero in the blue 
emission but nonzero in the red emission. An extra absorption of 
the secondary's light in the receding part of the disc causes 
the red emission to be reduced.
This effect should be balanced in the middle of the eclipse as seen
in Figure \ref{f6}. If this explanation is correct, the reverse
(stronger red emission) should be observed after the middle of the eclipse
at phases 0.5-0.6.
Unfortunately, these phases are not well covered by the observations
of Richards et al. (2005) but Figure 4 of Peters (1989) confirms that 
this is really the case. At the same time during secondary eclipse both 
emission peaks grow slightly.
This is mainly because the stellar continuum drops during the eclipse and 
the contribution of the emission to the total light can get more pronounced.
This is also reproduced in the synthetic spectra but the observed
emission peaks are still stronger than the theoretical ones which indicates 
that we may have underestimated the temperature of the secondary at this
phase or overestimated the adopted limb darkening of the secondary.
This might also be due to the reflection effect which is not considered 
in our calculations.

The blue emission extends to higher radial velocities (Figure \ref{f4}).
This is not predicted by the synthetic spectra and our interpretation
is that this is evidence of the gas stream. At this phase in 
the blue emission we are looking through 
more inhomogeneous regions encompassing the gas stream and disc-stream
interaction regions which have higher than average radial velocity
and can be rather turbulent exhibiting the strongest departures from 
circular Keplerian velocities. This is probably also the reason
why the observed central depression is displaced slightly to shorter 
wavelengths.
The observed central depression gets very deep at these phases,
and reaches a minimum at about the middle of secondary eclipse.
It is deeper than in the theoretical spectra. This huge depression 
could be caused by the combination of several effects:
(1) at this phase the intrinsic $H\alpha$ lines of both stars overlap
which deepens the central depression, 
(2) the reduction of the emission peaks by
an additional absorption of light from the secondary by the disc
might deepen the depression as well,
(3) as pointed out above, if the surface of the secondary were hotter 
at this phase it would make the intrinsic line of the secondary, as well
as the overall central depression, deeper;
(4) the eclipse of the secondary lowers the stellar continuum and changes 
both emission and absorption.

Something unexpected happens after the eclipse during phases
0.6-0.7. The blue emission increases again. This does not seem to be 
a temporal effect as it can be seen in the earlier observations
of Peters (1989; Figure 4). One could expect to observe 
the reverse from the opposite line of sight within phases 0.1-0.2
but it is not the case.
Could there exist some other circumstellar material apart from the disc
and the stream in the system that contributes to the emission? 
If this source were projected onto the surface of the secondary at phases
0.6-0.7 it might have the same effect as the disc at phases 0.4-0.5 
and might increase the relative strength of the blue emission.
If this were the case, on symmetry grounds, we could expect 
the reverse (red emission stronger) at phases 0.3-0.4.
There is only a very weak indication in Figure \ref{f6} that 
this might be the case. Independent data of Peters (1989) indicate 
the same effect that the red emission is slightly stronger at phases 
0.3-0.4.
However, if this matter projects onto the secondary during 0.6-0.7
and 0.3-0.4 then it would have to be outside of the Roche lobe of 
the primary. This may most probably be somewhere between the C1 and C2 
Roche surfaces.

\subsubsection{Second Quadrature}
The emission at the other quadrature (Figure \ref{f4}: phase 0.790) 
looks similar to that at 0.217 and is well reproduced in the calculations 
but now the observed central depression is weaker. 
The central absorption is smallest shortly after the quadrature 
at about 0.79-0.86 (Figure \ref{f6}). Our interpretation is that 
this is evidence of the hot spot or hotter disc regions projecting 
onto the stellar surface of the primary which increases the source 
function along the line of sight and reduces the central depression. 
A little bump seen now on the blue
side of the central depression is the intrinsic secondary $H\alpha$ line.

\subsubsection{Eclipse of the Disc}
The next four phases depicted in Figure \ref{f4} (phases 0.926-0.944) 
represent the eclipse of 
the approaching part of the disc and one can follow in both observed 
and theoretical spectra how the blue emission progressively ceases. 
The behavior of both emission peaks is thus explained fairly well with 
our model. Red emission seems to be stretching more to the red than 
predicted but this is just the manifestation of the same effect observed 
close to the secondary eclipse for the blue emission. This time, before
the primary eclipse in the red emission we observe the gas stream
with higher receding velocities than expected from circular Keplerian
orbits. However, the observed central absorption is again deeper 
than expected. Total eclipse which follows is discussed on a separate 
figure (Figure \ref{f5}) because of the different intensity scale.

The last four spectra in Figure \ref{f4} (phases 0.066-0.085) cover egress 
out of the disc eclipse. This time, one can observe how the red
emission associated with the receding part of the disc emerges.
Again, the behavior of emission peaks is reproduced fairly well in the
calculations, however, the observed central depression is much
deeper than expected from our model and poses the most serious problem. 
To some extent it may be caused by the similar effect invoked for the
explanation of the depression near secondary eclipse i.e., 
(a) overlap in radial velocities of stellar components 
(which operates only outside of the total eclipse this time).
We also encounter (b) the eclipse of the low velocity emission regions 
that spill into the central absorption which deepens the absorption. 
This is apparently not enough.
There are several possible explanations
of this additional absorption just before and after the primary eclipse.
It can be ascribed (1) to the main body of the disc and assumed that
the cooler disc regions project onto the hot surface of the primary slightly
before but, mainly, after the total eclipse. We reject this idea since such
cooler regions in the main body of the disc should manifest in asymmetric
emission peaks at the quadratures which is not observed.
Another alternative is (2) to ascribe these deep $H\alpha$ cores to 
the inner and more dense regions of the disc as suggested by Peters (1989). 
This might have no significant effect on the emission peaks but there
is another problem with this interpretation and it is the strong
phase dependence of this feature. There is no obvious reason why
this absorption if originating from inner disc regions should be phase 
dependent. One could assume that an inner disc region close to a hot spot
or a direct impact of the gas stream is hotter, but the matter would 
quickly cool towards the equilibrium value on time-scales much shorter 
than a Keplerian period of the inner disc. Moreover, we observe
that an enhanced absorption in the cores of $H\alpha$ are observed only 
shortly before and after the primary eclipse and quickly change and 
disappear within 0.1 orbital period between phases 0.85-0.95 and 
0.05-0.15. Such an enhanced absorption is observed at the same phases
in other Algols e.g., $\beta$~Per (Richards 1993) that does not have
a permanent accretion disc. 
This leads us to suggest a third possibility (3) that this effect could
be associated with some cooler matter in the vicinity of the secondary
star or L2 point. One could speculate that it may have the form of 
a slightly asymmetric envelope or a ``tail'' accompanying the secondary 
during its orbital motion or perhaps circumstellar matter populating
the region between the C1 and C2 critical Roche surfaces. 
In either situation, it is plausible that such matter could be present 
between C1 and C2, and could get there easily since the secondary 
already fills the C1 surface and is a magnetically active star. 
The disc itself reaches almost up to the Roche lobe of the primary and 
there is no reason to believe why it should have an abrupt end just 
before the Roche surface and not spill beyond the surface.  
The leading region (in the sense of the orbital movement) of such 
a secondary's envelope should be slightly hotter or less dense than 
the rear or tail of the envelope since the excess absorption is 
deeper after primary eclipse. 

There is still more to understand. Note, that just before
primary eclipse starts (0.90-0.95) the red emission is enhanced while
the blue one is being eclipsed. The opposite, even more pronounced
effect is observed after the eclipse (0.05-0.10) when the red emission
emerges out of eclipse as the blue emission gets fainter. 
It is not clear whether these features are permanent or transient
or why the non-eclipsed emission should be affected during the eclipse
of the other emission.
Possible explanations include (1) that the disc itself contributes 
significantly to the total continuum and this contribution is reduced 
during the disc eclipse or (2) the primary is being shadowed by the extra 
matter between C1 and C2 (as suggested above).
We made a test calculation for phase 0.0 with and without the disc
and found that the contribution of the uneclipsed disc to the whole light 
in the continuum at $H\alpha$ at this phase was only about $5\times 10^{-4}$.
Thus, the second option would be more plausible.
If the effect is real and the explanation correct one could expect
a drop in the R-band photometry light curve at the corresponding phases 
by about 0.05 mag.

\subsubsection{Total Eclipse}
Finally, Figure \ref{f5} exhibits the spectra during total eclipse.
This is an important verification of the disc model since
at this phase the main source of light, the primary star as well as
the inner disc regions, are obscured. The spectral lines of the secondary 
star can be seen more clearly and they indicate that the temperature
of the secondary is probably slightly higher than the value of 4361K
found by Van Hamme \& Wilson (1993). We adopted a temperature of 4600K
at the rotation pole of the secondary which is reduced by about 200K
on the back side of the secondary due to gravity darkening.
These phases also demonstrate that a slight decrease in radial density
profile of the disc is necessary to get better agreement with 
the observations.
This radial density profile is described with the power law behavior
$\rho=\rho_{in}(r/r_{in})^{\eta}$ in our model and the exponent 
$\eta=-1$ seems to fit the data best.
Both emission peaks agree fairly well with the synthetic
spectra. It is interesting that the red emission looks slightly
wider than the blue one. This is the manifestation of 
the same effect described above at the phases just before primary
eclipse or of the reversed effect seen close to secondary eclipse
when the blue emission was wider because of the enhanced departures 
from a Keplerian velocity field due to the gas stream and stream-disc 
interaction. However, at total eclipse the contribution from 
the remote part of the disc
close to the Roche lobe and out of the disc plane are mainly seen. 
These areas may exhibit the strongest departures from our simple 
prescription of the Keplerian velocity (single star model) 
which might explain why the synthetic central depression is wider.

The calculated central depression during primary eclipse is slightly 
stronger than expected. 
This can be due to several reasons: Firstly, the red 
emission which is wider overlaps the absorption and also skews it toward
the blue. Secondly, the synthetic spectra were calculated for 
the temperature of 4600K which may be attained at the rotational pole 
but the actual temperatures we observe at the back side of the secondary 
are lower by about 200K and the intrinsic Balmer lines of the secondary 
would be noticeably weaker at such temperatures.
We would like to note at this point that the skew of the central
depression to the blue at these phases (close to the primary and secondary
eclipse) can be easily misinterpreted as evidence of mass outflow.
At the other phases, like quadratures, departures from the circular 
Keplerian orbit may occur causing a shift of the central absorption 
to either side. That is why we feel that such blueshifted central 
depression is not proof of the mass outflow from the disc and 
that this conclusion of Peters (1989) should be revisited.

\subsubsection{Reliability of the free parameters of the disc}
Since there are so many free parameters of the disc we cannot claim 
that our fit is a unique solution. Nevertheless, as shown in the previous
section, the parameters affect the spectrum in specific ways that 
we now understand more clearly (see Figures \ref{f1}-\ref{f3}).
Consequently, such calculations can be used to put an independent constraint 
on many of these parameters. The calculations are not very sensitive to 
the inner radius of the disc but they seem to indicate that it reaches 
the surface of the star, moreover, the density of the disc seems 
to increase towards the star. This is expressed by the negative exponent 
$\eta$. However, the exponent is best constrained from the emission 
strength during the total eclipse which is also sensitive to 
the temperature of the secondary and if the latter is wrong  
a spurious value of $\eta$ could be obtained. If we had adopted
the cooler temperature of the secondary derived by 
Van Hamme \& Wilson (1993), the effect would be to lower the continuum 
and make the emission peaks relatively stronger and 
the exponent $\eta$ would have to be even more negative.
The outer radius can be set much more reliably since it affects 
strongly the separation of the emission peaks. The Roche lobe radius 
of the primary is about 10.6$R_{\odot}$ so our value is realistic and 
is about 95\% of the Roche lobe. This is in perfect agreement with 
the value suggested by Peters (1989) who estimated that the disc spans 
95\% of the Roche lobe. The exact agreement is a coincidence since 
in practice the disc probably does not have an abrupt end but rather 
dissolves smoothly.

The vertical dimension of the disc is significant and is of the order of 
the dimension of the primary star. It agrees with the estimates
of Peters (1989) and Plavec (1988) since it is in the region where
both their estimates overlap.
The temperature of the disc which we obtained should be viewed as
a characteristic temperature of the $H\alpha$ emitting 
regions. In reality, there may be strong local departures from such 
a homogeneous model.
Regions with temperatures between 5000K to 30000K will contribute 
significantly to the Balmer line emission but both cooler and hotter
regions may manifest at different wavelengths or spectral lines 
and an image of the disc can change depending on the 
spectral feature studied. The temperature was inferred mainly from 
the central depression which is strongly variable and there may be 
other effects which are beyond the scope of the paper
and may turn out to be important like NLTE effects, a better 
treatment of line absorption coefficient and scattering at optically 
thick wavelengths.

The densities are quite well determined since the overall emission
is very sensitive to this parameter. Nevertheless, in reality, 
again strong local departures can be expected from such a simple density 
profile. The densities we obtained throughout the disc decline
from $n_{e}=2\times 10^{10}$cm$^{-3}$ at the inner radius to 
$n_{e}=4\times 10^{9}$cm$^{-3}$ at the outer radius. They are in very good 
agreement with the estimate of Peters (1989; $10^{10}$cm$^{-3}$).
The assumed microturbulence is very uncertain. In the case of an optically
thin disc, it should be viewed as departures from the Keplerian velocity 
field rather than a local turbulence, and we can claim only that 
these departures will not be higher than about 80 km\,s$^{-1}$. 
However, these departures may be higher in the gas stream.
Consequently, we can claim that the single star disc model with 
a Keplerian velocity field (as given by Eq. \ref{e1}) can reproduce 
the shape of the line fairly well. The regions of the disc out of 
the orbital plane and close to the Roche lobe are most vulnerable 
to departures from this model. 

The Voigt profile is generally not a good approximation for hydrogen 
lines, however, it can be justified in the case of emission lines 
originating from low density regions where the line profile is 
determined mainly by the 3D velocity field.
Finally, we would like to point out that the parallax measured by Hipparcos 
corresponds to the distance of about $154\pm 25$pc which is not 
in very good agreement with the previous estimates of Etzel (1988) 
and Plavec (1988) who both obtained distances of about 194 pc. 
If the new distance is confirmed by future parallax 
measurements, then either the geometrical scale (semi-major axis) and 
stellar radii are smaller than those obtained by Etzel(1988), Plavec (1988)
or Van Hamme \& Wilson (1993) or the temperatures of the stars, 
mainly that of the primary, might have to be reduced slightly.

%\clearpage
%\clearpage

\section{Summary and Conclusions}
We have developed a new computer code called {\sc shellspec}.
This is to our knowledge the only tool which can produce the synthetic light
curves and synthetic spectra of interacting binaries taking into account
optically thin moving circumstellar matter. The code was applied to
the well-known eclipsing Algol-type binary system TT Hya which has 
an accretion disc. Synthetic spectra of the system for all phases were 
calculated by taking into account spherical primary, a Roche lobe filling 
secondary,  as well as a disc.
For the first time, it has been verified by calculations of synthetic spectra 
that the double peaked emission in TT Hya, and thus also 
in other long period Algol-type eclipsing binaries, 
originates from circumstellar matter in the form of an accretion disc 
with a Keplerian velocity field.

The effect on the spectrum of various free 
parameters was studied. The temperature and inclination
of the disc have the strongest effect on the depth of the central depression
while the outer radius of the disc, the radial density profile,
and the inclination affect mainly the position and separation 
of the emission peaks. The overall strength of the emission is regulated
mainly by the density and temperature.
The central depression was found to be created by two different effects:
the geometry and the velocity field of the disc itself and
by the cool matter projected onto the hotter stellar surface.

Realistic estimates of the densities, temperature, geometry, and 
dynamics of the disc were obtained as a result of these calculations.
However, there are differences between the observed and synthetic
spectra which assume a simple circular Keplerian disc.
These can be attributed mainly to the gas stream or stream-disc interaction.
The presence of an additional source of circumstellar matter in the system
between the C1 and C2 Roche surfaces is suggested to account for the 
deep absorption in the $H\alpha$ cores detected near primary eclipse
and various other observed features.

\acknowledgements
We would like to thank Drs. Ivan Hubeny, Alon Retter, and the referee, 
for their comments and suggestions, and Dr. Pavel Koubsk\'{y} for kindly 
allowing us to use our jointly collected 1996/1997 spectra of TT Hya
prior to publication.
JB gratefully acknowledges grant support from Penn State University
and thanks Dr. Konstatin Getman, and department computer staff for 
their assistance with computer related problems.
This research was supported by the NSF-NATO Fellowship (NSF DGE-0312144)
and partly by the VEGA grant No. 3014 from the Slovak Academy 
of Sciences and the Science and Technology Assistance agency under
the contract No. 51-000802.
%This study made use of the Vienna Atomic Line Data Base (VALD) services.

\end{document}